\documentclass[12pt]{iopart}

\expandafter\let\csname equation*\endcsname\relax
\expandafter\let\csname endequation*\endcsname\relax
\usepackage{caption}
\usepackage[usenames,dvipsnames]{color}
\usepackage[utf8]{inputenc}
\usepackage{lipsum}
\usepackage[sc]{mathpazo}
\usepackage{graphicx}
\usepackage{amssymb}
\usepackage{subfigure}
\usepackage{stackrel}
\usepackage{enumitem}
\usepackage{amsmath}
\usepackage{amsfonts}
\usepackage{bm}
\usepackage{color}
\usepackage{verbatim} 
\usepackage{tcolorbox}
\usepackage[normalem]{ulem}
\usepackage{hyperref}
\usepackage{soul}

\usepackage[usenames,dvipsnames]{color}
\usepackage{amsfonts}
\usepackage[noadjust]{cite}

\newlist{subquestion}{enumerate}{1}
\setlist[subquestion,1]{label=(\alph*)}

\newcommand{\ket}[1]{\ensuremath{\left|{#1}\right\rangle}}
\newcommand{\bra}[1]{\ensuremath{\left\langle{#1}\right |}}

\newcommand{\ojo}[1]{#1}

\newcommand{\beq}{\begin{equation}}
\newcommand{\eeq}{\end{equation}}
\newcommand{\bse}{\begin{subequations}}
	\newcommand{\ese}{\end{subequations}}\newcommand{\bea}{\begin{eqnarray}}
\newcommand{\eea}{\end{eqnarray}}
\newcommand{\bit}{\begin{itemize}}
	\newcommand{\eit}{\end{itemize}}
\newcommand{\bpmatrix}{\begin{pmatrix}}
	\newcommand{\epmatrix}{\end{pmatrix}}

\newcommand{\be}{\begin{equation}}
\newcommand{\ee}{\end{equation}}
\newcommand{\ben}{\begin{eqnarray}}
\newcommand{\een}{\end{eqnarray}}

\begin{document}
\title{Persistent fermionic entanglement under decoherence}

\author{E. Salinas$^{1}$, A. P. Majtey$^{2}$ and A. Vald\'{e}s-Hern\'{a}ndez$^1$}
\address{$^1$Instituto de F\'{\i}sica, Universidad Nacional Aut\'{o}noma de M\'{e}xico, Apartado Postal 20-364, Ciudad de M\'{e}xico, Mexico}
\address{$^2$Instituto de F\'{i}sica Enrique Gaviola, CONICET and Universidad 
Nacional de C\'{o}rdoba,
Ciudad Universitaria, X5016LAE, C\'{o}rdoba, Argentina}
\ead{andreavh@fisica.unam.mx}

\begin{abstract}
 We consider a system of two indistinguishable fermions (with four accessible states each) that suffers decoherence without dissipation due to its coupling with a global bosonic bath at a fixed temperature. 
Using an appropriate measure of fermionic entanglement, we identify families of two-fermion states  whose entanglement persists throughout the evolution, either fully or partially, despite the noisy effects of the interaction with the bath, and independently of its temperature.
The identified resilience to decoherence provides valuable insights into the entanglement dynamics of open systems of indistinguishable fermions, and into the conditions under which long-lived entanglement emerges under more general decoherence channels. 
\end{abstract}

\noindent{\it Keywords}: \ojo{Decoherence, Fermionic system, Persistent entanglement}

 
\section{Introduction}

Identical-particle systems constitute a promising quantum platform   
for information processing \cite{LoFranco_2018,Morris_2020}. 
At the heart of this promise lies the concept of indistinguishability, which endows composites of identical parties with highly non-classical features. 
While operational frameworks on distinguishable-particle systems are well-established through individual operations on each party, the utilization of indistinguishable particles, which are inherently individually unaddressable, poses a persistent challenge, particularly pronounced in the context of quantum technologies. 

Entanglement, a crucial resource for many quantum protocols in metrology and quantum information processing, is well understood for systems of distinguishable particles \cite{nielsen_chuang_book,amico_2008,horodecki_2009}. 
For indistinguishable parties advances have been made \cite{Ghirardi_2002,Eckert_2002,tichy_2011_JPB,Benatti_2020}, yet the lack of a general consensus regarding the very definition of entanglement in such systems has prompted the introduction of diverse approaches to tackle  
its quantification, and the feasibility of obtaining useful (accessible) entanglement from a composite of  indistinguishable particles \cite{killoran_2014,bouvrie_2017_b,bouvrie_2019,Jimenez2024}.
In this work, we stick to the approach according to which a separable, i.e. non-entangled, pure state of two indistinguishable fermions is one that can be written as a single Slater determinant \cite{Ghirardi_2002,Eckert_2002,ghirardi_2004,Majtey2023}.  

The system under study is a pair of non-interacting identical fermions that interact globally with a bosonic bath at temperature $T$, under a non-dissipative and decoherence channel that generalizes the well-known spin-boson model \cite{Legget1987,Weiss2012} to describe a generic quantum system embedded in a collection of harmonic oscillators \cite{Privman1998}. 
In \cite{Bussandri2020} this system was considered, and three qualitatively different types of entanglement dynamics were identified, depending on the initial two-fermion state: 
Invariant regime (for states that pertain to the decoherence-free subspace), exponential decay of entanglement, and sudden death of entanglement (sudden vanishing of the entanglement  while  coherence decays exponentially).
Here we contribute to those results with the identification of additional subspaces in which the pair of fermions 
exhibit persistent non-zero entanglement, resilient to the effect of decoherence. This finding does not only reinforce the richness of the model, but also reveals that long-lived fermionic entanglement can be attained under decoherence dynamics, provided appropriate initial conditions are met.

The phenomenon of persistent correlations enables more reliable and efficient quantum information tasks, and  
has been studied previously in different systems. 
For example, Ref. \cite{Dajka2007} considers a system of two non-interacting qubits, one of which is non-dissipatively but dephasingly coupled to the environment, and finds that under certain regimes
entanglement can persist for a long time, leading the authors to conjecture that certain composite systems could remain entangled indefinitely, even when coupled to a thermal bath. 
In \cite{Yu2006}, the effects of classical noise on the dynamics of two-qubit entanglement was discussed,  highlighting the existence of long-lived entanglement in the presence of global dephasing noise.
Further, the influence of white
noise on the entanglement dynamics of two atoms in two cavities has been investigated in \cite{Xu2005}. 
The study found that when a single atom is driven by a white-noise field, it results in long-lived entanglement.
It has also been shown that a phase-amplified cavity can lead to persistent entanglement in a quantum system consisting of a single Cooper-pair box irradiated by a quantized field \cite{Abdel-Aty2008}. 
In \cite{Zhang2010} it was pointed out that in the long-term, the entanglement between a pair of two-level atoms interacting with two coherent fields of two spatially separated and dissipative cavities, tends to a fixed value under certain conditions.  
%
In the context of fermionic environments, the development of exact Grassmann stochastic Schr\"odinger equations and non-Markovian quantum state diffusion methods allow for the study of open fermionic systems, providing insights into quantum decoherence and transport processes \cite{PRA871,PRA872}. In particular, these works show that quantum coherence can be deeply modified by environmental memory effects, a feature that may impact the entanglement in the long-time regime.

The present paper complements these and related works limited to composites of distinguishable parties, by focusing on the emergence of persistent entanglement in an open system of indistinguishable fermions. 
In particular, we identify families of entangled states that evolve into states that preserve (completely or partially) the amount of entanglement between the fermions, despite the detrimental effects of decoherence induced by the bath. 
This complementary perspective enhances our understanding of the conditions under which robust entanglement can be achieved, and particularly offers a strategy, based on the preparation states, for advancing the practical applications of quantum information science in fermionic entangled systems.

The article is structured as follows. In Section 2 we introduce the basic framework for quantifying 
the entanglement and the coherence in the indistinguishable-fermion system. 
Section 3 is devoted to presenting the surrounding bath and the dynamical model under which the two-fermion system will evolve, along with a brief description of some previously identified dynamics of the two-fermion entanglement. 
Our main results are presented in Section 4.
There we compute the coherence and the fermion-fermion entanglement in the long-time regime, and establish the conditions (on the initially pure two-fermion state) for having persistent entanglement. 
Families of states with invariant (non-evolving) entanglement are identified, and a numerical analysis is performed to characterize more general classes of states that  
evolve towards ones with persistent entanglement.
Further, an analysis of the persistence factor is made, thus allowing for the characterization of the initial fermionic states according to the percentage of entanglement that endures in time.
Finally, we briefly comment on the extension of this analysis to other decoherence processes, taking as an example the amplitude decay as a paradigmatic dissipative quantum channel. 
We summarize and conclude this work in Section 5. 


\section{Two-fermion entangled system} 

We consider a system composed of two indistinguishable fermions, each of which has $d=4$ accessible orthogonal states $\{ \ket{i} \} = \{\ket{1},\ket{2},\ket{3},\ket{4} \}$ that define a basis of the single-particle Hilbert space $\mathcal{H}_f$.
The antisymmetric Hilbert space  $\mathcal{H_-}=\mathcal{H}_f\land\mathcal{H}_f$ is thus a $d(d-1)/2=6$-dimensional space, spanned by vectors $\{ \ket{\psi_n^-} \}$ with $n\in \{ 1,2,...,6 \}$.
%

In the angular momentum representation, the basis $\{ \ket{\psi_n^-} \}$ is identified with the antisymmetric eigenstates $\{\ket{j,m} \}$ of the total angular momentum operators $\boldsymbol{J}^2$ and $J_z$, with $0\leq j\leq 2s$ and $s$ determined according to $d=2s+1$. The basis $\{\ket{i}\}$ is identified with the eigenstates $\{\ket{s,m_s} \}$ of the single-particle angular momentum operators $\boldsymbol{J}_1^2$ and $J_{z1}$. 
In the present case with $d=4$, we have $j=0,1,2,3$, yet only for $j=0,2$ the eigenvectors $\{\ket{j,m}\}$ are antisymmetric.
The basis of $\mathcal H_-$ and $\mathcal H_f$ in this representation thus read, respectively,
\begin{equation}\label{b1}
\left\{\left|\psi_{n}^{-}\right\rangle\right\}=\{|2,2\rangle,|2,1\rangle,|2,0\rangle,|2,-1\rangle,|2,-2\rangle,|0,0\rangle\},
\end{equation}
and
\begin{equation}\label{b2}
\{\ket{i}\}=\{
\ket{\tfrac{3}{2},\tfrac{3}{2}},
\ket{\tfrac{3}{2},\tfrac{1}{2}},
\ket{\tfrac{3}{2},-\tfrac{1}{2}},
\ket{\tfrac{3}{2},-\tfrac{3}{2}}
\}.
\end{equation}

The Clebsh-Gordan coefficients allow us to decompose the vectors $\{\ket{\psi_{n}^{-}}\}$ in the uncoupled basis $\{\ket{i}\otimes\ket{j}=\ket{i}\ket{j}\}$, leading to 
\begin{subequations}\label{tablaedos}
\begin{eqnarray}
|\psi_{1}^{-}\rangle&=&|2,2\rangle=\left|\psi_{12}^{sl}\right\rangle,\\
|\psi_{2}^{-}\rangle&=&|2,1\rangle=\left|\psi_{13}^{sl}\right\rangle,\\
|\psi_{3}^{-}\rangle&=&|2,0\rangle=\tfrac{1}{\sqrt{2}}\left(\left|\psi_{14}^{sl}\right\rangle+\left|\psi_{23}^{sl}\right\rangle\right),\\
|\psi_{4}^{-}\rangle&=&|2,-1\rangle=\left|\psi_{24}^{sl}\right\rangle,\\
|\psi_{5}^{-}\rangle&=&|2,-2\rangle=\left|\psi_{34}^{sl}\right\rangle,\\
|\psi_{6}^{-}\rangle&=&|0,0\rangle=\tfrac{1}{\sqrt{2}}\left(\left|\psi_{14}^{sl}\right\rangle-\left|\psi_{23}^{sl}\right\rangle\right),
\end{eqnarray}
\end{subequations}
where $\{\ket{\psi_{12}^{sl}},\ket{\psi_{13}^{sl}},\ket{\psi_{14}^{sl}},\ket{\psi_{24}^{sl}},\ket{\psi_{34}^{sl}},\ket{\psi_{23}^{sl}}\}$ defines the (ordered) basis of Slater determinants (that span $\mathcal H_-$)
\begin{equation}
\left|\psi_{i j}^{sl}\right\rangle\equiv
\tfrac{1}{\sqrt{2}}(\ket{i}\ket{j}-\ket{j}\ket{i}), \quad(i \neq j).    
\end{equation}
The unitary transformation that takes the basis $\{|\psi_{n}^{-}\rangle\}$ into $\{|\psi_{ij}^{sl}\rangle\}$ is thus
\begin{equation}\label{U}
U=\left(\begin{array}{cccccc}
1 & 0 & 0 & 0 & 0 & 0 \\
0 & 1 & 0 & 0 & 0 & 0 \\
0 & 0 & \tfrac{1}{\sqrt{2}} & 0 & 0 & \tfrac{1}{\sqrt{2}} \\
0 & 0 & 0 & 1 & 0 & 0 \\
0 & 0 & 0 & 0 & 1 & 0 \\
0 & 0 & \tfrac{1}{\sqrt{2}} & 0 & 0 & -\tfrac{1}{\sqrt{2}}
\end{array}\right).
\end{equation}

An arbitrary pure state of the two-fermion system can therefore be decomposed as 

\begin{equation}\label{psi_init}
   \ket{\psi_0} = \sum_{n=1}^{6} \alpha_n \ket{\psi^-_n}=\sum^4_{i,j=1 }w_{ij}\ket{\psi_{ij}^{sl}},
\end{equation}
with $w_{ij}=-w_{ji}$, and $\sum_n|\alpha_n|^2=4\sum_{i,j}|w_{ij}|^2=1$. The coefficients of each expansion are related according to:
\begin{subequations}
\begin{eqnarray}
w_{12}&=&\tfrac{1}{2}\alpha_1,\quad w_{13}=\tfrac{1}{2}\alpha_2, \quad w_{24}=\tfrac{1}{2}\alpha_4
,\quad w_{34}=\tfrac{1}{2}\alpha_5,\\
w_{14}&=&\tfrac{1}{2\sqrt{2}}(\alpha_3+\alpha_6),\quad
w_{23}=\tfrac{1}{2\sqrt{2}}(\alpha_3-\alpha_6).
\end{eqnarray}
\end{subequations}

\subsection{Entanglement and coherence in the two-fermion system}
Throughout the paper we will stick to the definition of entanglement in composites of indistinguishable fermions according to which a state of the two-fermion system is separable, or non-entangled, if and only if the corresponding density matrix can be decomposed into a convex sum of pure states with Slater rank 1 in some basis:
\begin{equation}
\rho^{\mathrm{sep}}=\sum_{\alpha ,\beta} p_{\alpha \beta}\left|\psi_{\alpha\beta}^{sl}\right\rangle\left\langle\psi_{\alpha\beta}^{sl}\right|,
\end{equation}
where $p_{\alpha\beta} \geqslant 0$,  $\sum_{\alpha,\beta} p_{\alpha\beta}=1$. 
Otherwise the pair of fermions will be said to be entangled \cite{Ghirardi_2002,Eckert_2002}.  
In particular, a two-indistinguishable-fermion pure state is regarded as separable (non-entangled) if and only if it can be expressed as a single Slater determinant. In such case the only correlations present are the exchange correlations, originated in the antisymmetrization of the state vector. 
This evinces that the present notion of fermionic entanglement refers to those correlations on top of the exchange correlations. 
For $d=4$ the amount of {\it fermionic} entanglement in a two-fermion state $\rho$
can be quantified by the {\it fermionic concurrence} $C_f(\rho)$, defined as \cite{Eckert_2002}
\begin{equation}\label{Cf}
C_{f}\left(\rho\right)=\max \left\{0, \lambda_{1}-\lambda_{2}-\lambda_{3}-\lambda_{4}-\lambda_{5}-\lambda_{6}\right\},
\end{equation}
where $\{\lambda_i\}$ are the square roots of the eigenvalues, arranged in decreasing order, of the matrix $\rho\mathbb{D} \rho \mathbb{D}^{-1}$, where
\begin{equation}\label{matrizD}
\mathbb{D}=\left(\begin{array}{cccccc}
0 & 0 & 0 & 0 & 1 & 0 \\
0 & 0 & 0 & -1 & 0 & 0 \\
0 & 0 & 1 & 0 & 0 & 0 \\
0 & -1 & 0 & 0 & 0 & 0 \\
1 & 0 & 0 & 0 & 0 & 0 \\
0 & 0 & 0 & 0 & 0 & 1
\end{array}\right) \kappa,
\end{equation}
%
$\kappa$ stands for the complex conjugation operator, and the  
matrix is expressed in the basis $\mathcal B=\{|2,2\rangle,|2,1\rangle,|2,0\rangle,|2,-1\rangle,|2,-2\rangle,i|0,0\rangle \}$ (notice that it differs from the basis $\{|\psi^-_n\rangle\}$ by the imaginary unit in front of $|\psi^-_6\rangle$). 
In the pure state case, the expression for the concurrence (\ref{Cf}) simplifies as \cite{Eckert_2002}:
\begin{equation}\label{concurrencef}
 C_f(\ket{\psi_0})=8|w_{12}w_{34}-w_{13}w_{24}+w_{14}w_{23}|
 =|\alpha^2_3-\alpha^2_6+2(\alpha_1\alpha_5-\alpha_2\alpha_4)|,
\end{equation}
which is bounded as $0\leq C_f\leq 1$.

Coherence is an extremely useful quantum resource. It lies at the root of the most distinctive quantum phenomena, and in particular allows for the existence of quantum correlations, specifically of entanglement \cite{Adesso2016}.
However, unlike the latter, the coherence is a basis-dependent quantity. 
A legitimate measure of the coherence $\mathcal K$ of an arbitrary density matrix $\rho$ in a given basis $\{\ket{\mu}\}$ is given by \cite{Plenio2014}
\begin{equation}
   \mathcal K =  \sum_{\nu\neq \mu} |\rho_{\nu\mu}| 
   \label{C}
\end{equation}
with $\rho_{\nu\mu}=\langle \nu|\rho|\mu\rangle$.  

According to the definition of fermionic entanglement presented above, the existence of entanglement in identical-fermion systems relies on the coherent superposition of Slater determinants. In line with this approach, the relevant basis for quantifying the coherence is $\{|\psi_{i j}^{sl}\rangle\}$. 
Resorting to this preferred basis thus guarantees that the absence of coherence implies the absence of entanglement, or equivalently, that states with non-zero amount of entanglement correspond to coherent states in the Slater basis.
Throughout the paper we will thus quantify $\mathcal K$ in the basis $\{|\psi_{i j}^{sl}\rangle\}$.

\section{Dynamical Model}\label{dinmodel}

Following \cite{Privman1998,Bussandri2020}, we consider a pair $(S)$ of non-interacting and indistinguishable fermions coupled to a bosonic bath $(E)$ under a non-dissipative interaction. 
The complete Hamiltonian writes as (we take $\hbar=1$): 
\begin{equation}\label{Hamil}
H=H_{S}+\sum_{k} \omega_{k} a_{k}^{\dagger} a_{k}+\Lambda_{S} \sum_{k}\left(g_{k}^{*} a_{k}+g_{k} a_{k}^{\dagger}\right).    
\end{equation}
The operators $a_{k}^{\dagger}$ and $a_{k}$  stand for the (bosonic) creation and annihilation operators, respectively, corresponding to the $k$-th mode of the bosonic bath, with frequency $\omega_k$. 
The scalar $g_{k}$ is a coupling constant, and $\Lambda_{S}$ an observable of $S$ satisfying 
$\left[H_{S}, \Lambda_{S}\right]=0$, in line with the non-dissipative condition $[H_S,H]=0$.
The Hamiltonian (\ref{Hamil}) generalizes the simplest version of the spin-boson model ---which from among the various models of decoherence stands out as a paradigmatic and versatile one---, to linearly couple an {\it arbitrary}-dimensional system $S$ to a collection of harmonic oscillators (a model for the environment that works particularly well in a wide range of situations \cite{reviewDeco}). In the present case, $S$ is made up of a pair of 4-dimensional fermions, collectively coupled to the bosonic environment. 
Our analysis thus focuses on a direct extension of the spin-boson model applied to a system that can be encountered, for example, in architectures involving two 1/2 spin particles in a double-well potential, in the emergent study of effective 3/2 spin particles \cite{BoettcherPRL2020}, or more generally in higher-spin systems coupled to the bosonic bath \cite{PRL2005}.
Further, this type of model, corresponding to decoherence without dissipation, is suitable for describing decoherence processes that occur on a time scale that is much shorter than the time scale for dissipation. This is of particular interest in a wide variety of situations aimed at preserving coherence without energy loss (as is desirable in precision measurements and quantum computing), and have been studied in optomechanical systems, cavity quantum electrodynamics, quantum metrology, etc. \cite{Bonifacio_2000,Banerjee2007,Chakrabarty2011}

The property $[H_S,H]=0$ defines a basis of $\mathcal H_S$ conformed by the common eigenvectors $\{\ket{n}\}$ of  
 $H_{S}$ and $\Lambda_{S}$, 
\begin{equation}\label{eigenH}
H_{S}|n\rangle=E_{n}|n\rangle, \quad \Lambda_{S}|n\rangle=L_{n}|n\rangle . 
\end{equation}

Assuming that $S$ and $E$ are initially uncorrelated, and that $E$ is initially in a thermal state at temperature $T$, the matrix elements of the evolved reduced density matrix $\rho_S(t)=\rho(t)$, in the basis $\{\ket{n}\}$ and in the interaction picture, read \cite{Privman1998,Bussandri2020}   
\begin{subequations}\label{rhomn}
\begin{equation}\label{rhoevol2}
    \langle m|\rho_S(t)|n\rangle\equiv\rho_{mn} (t) = \rho_{mn}(0)  f_{mn} (t), 
\end{equation}
with 
\begin{equation}\label{fmn}
f_{mn}(t)=e^{-(L_m-L_n)^2\Gamma(t)}e^{-i(L_m^2-L_n^2)r(t)},  
\end{equation}
\end{subequations}
where $r(t)=\Delta(t)-\Theta(t)$, and 
\begin{subequations}\label{funciones}
\begin{eqnarray}
    \Gamma(t)&=&2\sum_k\omega_k^{-2} |g_k|^2 \sin^2 \frac{\omega_k t}{2} \coth \frac{\beta\omega_k}{2},
    \\\label{28a}
    \Delta(t)&=&\sum_k\omega_k^{-2} |g_k|^2 \sin \omega_k t,
\\\label{28b}
    \Theta(t)&=&\sum_k\omega_k^{-2} |g_k|^2 \omega_k t.
\label{28c}
\end{eqnarray}
\end{subequations}

Equations \eqref{rhomn} reveal that, irrespective of the specificities of $H_S$ and $\Lambda_S$, the populations $\rho_{nn}$ are not affected by the presence of the bath, as expected for a non-dissipative interaction. Further, if there is degeneracy in $\Lambda_S$ ($L_m=L_n$ with $n\neq m$), then $f_{nm}(t)=1$ and the corresponding density matrix element does not evolve in time. The degeneracy in $\Lambda_S$ thus defines a Decoherence Free (DF) subspace, whose elements are states that remain invariant under the evolution.
Moreover, if also there is degeneracy in $\Lambda^2_S$ (so that $L^2_m=L^2_n$ with  $L_m\neq L_n$), then only the decaying term contributes to $f_{nm}(t)$; consequently, the degeneracy of $\Lambda^2_S$ defines Exponential Decay (ED) subspaces, in which the dynamics is characterized by an exponential decay.  
These observations indicate that by choosing a common eigenbasis  $\{\ket{n}\}$ in which both $\Lambda_S$ and $\Lambda^2_S$ exhibit degeneracy, allows for the exploration of varied and rich dynamics. 

Assuming further that the bath is large enough so the density of modes can be considered continuous, we can substitute the discrete sums in (\ref{funciones}) by integrals according to the prescription
$\sum_k |g_k|^2 \rightarrow \int_0^\infty d\omega J(\omega)$,
with $J(\omega)$ the spectral density. Here we take $ J(\omega)=4J_0\omega e^{-(\omega/\omega_c)}$, with $J_0$ a dimensionless constant and $\omega_c$ the cutoff frequency, which in turn defines  
the characteristic temperature $T_c = \omega_c$.
With these considerations, Eqs. (\ref{funciones}) become
\begin{subequations}\label{continuum}
\begin{eqnarray}
    \Gamma(t) &=& \frac{J_0}{2} \int_0^\infty d\omega \,e^{-(\omega/\omega_c)}   \,\frac{\sin^2 \frac{\omega t}{2}}{\omega} \coth \frac{\beta\omega}{2},
    \\\label{int1}
     \Delta(t) &=&  \int_0^\infty d\omega \, e^{-(\omega/\omega_c)} \frac{\sin \omega t}{\omega} ,
    \\\label{int2}
     \Theta(t) &=& t\int_0^\infty d\omega \, e^{-(\omega/\omega_c)} .
     \label{int3}
\end{eqnarray}
\end{subequations}
We can explicitly show that $\Gamma(t)$ is an increasing function of $t$ by making the change of variable $x=\omega t$, obtaining
\begin{eqnarray}\label{Gammachange}
\Gamma(t) &=& \frac{J_0}{2} \int_0^\infty dx \,e^{-(x/\omega_c t)}   \,\frac{\sin^2 \frac{x}{2}}{x} \coth \frac{\beta x}{2t}.
\end{eqnarray}
The functions $e^{-(x/\omega_c t)}$ and $\coth \frac{\beta x}{2t}$ are both positive increasing functions of $t$, so their product is also an increasing function of $t$. Consequently the integrand in (\ref{Gammachange}), and hence $\Gamma(t)$, increases with $t$, and the first factor in Eq. (\ref{fmn}) decreases exponentially in time. 


\subsection{Dynamics of entanglement}
In order to analyze the detailed dynamics of entanglement in the present model, the specification of $\Lambda_S$ and $H_S$ is required. In line with the observations below Eqs. (\ref{funciones}), diverse dynamics can be observed by choosing $\Lambda_S=J_z$ and identifying the common eigenbasis $\{\ket{n}\}$ with the set $\{\ket{\psi^{-}_n}\}$ of angular-momentum eigenvectors in Eq. (\ref{b1}). 
The fermionic Hamiltonian $H_S$ can thus be taken, for example, proportional to $J_z$ or to $\boldsymbol J^2$. Its specific election is unimportant for the present analysis, yet for $H_S\sim J_z$ contact is established with the higher-spin generalization of the spin-boson model (applicable, in particular, to our 6-dimensional fermionic composite, which can be mapped to a 5/2-spin system). 
By fixing $\Lambda_S=J_z$ and $\{\ket{n}\}=\{\ket{\psi^{-}_n}\}$, the eigenvalues of $\Lambda_S$  read 
\beq
L_1=-L_5=2;\;\;\;L_2=-L_4=1;\;\;\;L_3=L_6=0.
\eeq

We further assume that initially the fermions are prepared in a pure state $\ket{\psi_0}$, that generically writes as (\ref{psi_init}). 
As discussed  previously, qualitative different dynamics of the entanglement between the fermions may appear in this scenario, depending on $\ket{\psi_0}$ \cite{Bussandri2020}. 
The vectors $\ket{\psi_3^-}$ and $\ket{\psi_6^-}$ span the Decoherence Free  subspace, with elements
\beq
\label{DF}
\ket{\psi_0}_{\textrm{DF}} = \alpha_3\ket{\psi_3^-} + \alpha_6\ket{\psi_6^-},
\eeq
satisfying $\rho(t)=\rho(0)=\ket{\psi_0}\!\bra{\psi_0}_{\textrm{DF}}$. This implies, in particular, that the entanglement between the fermions is constant.
Also, two Exponential Decay  subspaces are identified, with states
\beq\label{expdecay}
    \ket{\psi_0}_\textrm{ED} =  \alpha_n\ket{\psi_n^-} + \alpha_m\ket{\psi_m^-} ,
 \eeq
with $(n,m)=(2,4), (1,5)$.
%
These states are characterized by an entanglement dynamics that decays exponentially in time.
%
For other initial states not belonging to the DF or the ED subspaces, the oscillatory terms in (\ref{fmn}) appear, and  interesting entanglement dynamics may arise, as for example, fermionic entanglement sudden death \cite{Bussandri2020}.
\section{Persistence of entanglement under decoherence}\label{persis}

The varied dynamics that can emerge in the fermionic system poses the question as to whether there exist subspaces whose elements $\ket{\psi_0}$ evolve into states whose entanglement endures in the long-time regime. 
%
To address this question  
we first explore the structure of the two-fermion density matrix 
$\rho$ as $t\rightarrow\infty$.

Since $\Gamma(t)$ is a positive and increasing function of $t$, it follows from 
 (\ref{rhomn})  that as $t\rightarrow\infty$ the only matrix elements that endure are those for which $L_m=L_n$, that is, the diagonal and the off-diagonal elements corresponding to degenerate states, $L_3=L_6$.
Therefore, the matrix $\rho(t\rightarrow\infty)$ in the basis 
 $\{\ket{\psi_n^-}\}$ is 
\begin{equation}\label{rhoTinft}
\rho(t\rightarrow\infty)=\left(\begin{array}{cccccc}
\rho_{11} & 0 & 0 & 0 & 0 & 0 \\
0 & \rho_{22} & 0 & 0 & 0 & 0 \\
0 & 0 & \rho_{33} & 0 & 0 & \rho_{36} \\
0 & 0 & 0 & \rho_{44} & 0 & 0 \\
0 & 0 & 0 & 0 & \rho_{55} & 0 \\
0 & 0 & \rho_{63} & 0 & 0 & \rho_{66}
\end{array}\right),
\end{equation}
whereas in the basis $\{\ket{\psi_{ij}^{sl}}\}$ it reads
%
\begin{equation}\label{rhoSlTinft}
\rho^{sl}(t\rightarrow\infty)=\left(\begin{array}{cccccc}
\rho_{11} & 0 & 0 & 0 & 0 & 0 \\
0 & \rho_{22} & 0 & 0 & 0 & 0 \\
0 & 0 & \frac{1}{2}(\rho_{33}+\rho_{66}+2\textrm{Re}\rho_{36}) & 0 & 0 & \frac{1}{2}(\rho_{33}-\rho_{66}-2i\textrm{Im}\rho_{36}) \\
0 & 0 & 0 & \rho_{44} & 0 & 0 \\
0 & 0 & 0 & 0 & \rho_{55} & 0 \\
0 & 0 & \frac{1}{2}(\rho_{33}-\rho_{66}+2i\textrm{Im}\rho_{36}) & 0 & 0 & \frac{1}{2}(\rho_{33}+\rho_{66}-2\textrm{Re}\rho_{36})
\end{array}\right).
\end{equation}
%
Since the matrix elements involved  are time-independent, it follows that in the expressions for the asymptotic state we have 
\beq
\rho_{nm}=\rho_{nm}(0)=\alpha_n\alpha^*_m.
\eeq

The coherence in the long-time regime, $\mathcal K_{\infty}\equiv\mathcal K(t\rightarrow\infty)$, thus reads
\beq \label{cohesl}
\mathcal K_{\infty}=\Big||\alpha_3|^2-|\alpha_6|^2+2\,i\,\textrm{Im}\,\alpha_3\alpha^*_6 \Big|.
\eeq
Consequently, a non-vanishing overlap of the initial state with the DF subspace suffices for having a persistent coherence (provided $\alpha_3\neq\alpha_6$), and thereby is a necessary condition for having persistent concurrence.
Writing $\alpha_3= |\alpha_3|e^{i\theta_3}$, $\alpha_6=|\alpha_6|e^{i\theta_6}$, and $\theta=\theta_3-\theta_6$, we get 
\beq \label{cohesl2}
\mathcal K^2_{\infty}=|\alpha_3|^4+|\alpha_6|^4-2|\alpha_3\alpha_6|^2\cos2\theta,
\eeq 
whence 
\beq\label{boundcoherence}
\Big||\alpha_3|^2-|\alpha_6|^2\Big|\leq\mathcal K_{\infty}\leq \Big||\alpha_3|^2+|\alpha_6|^2\Big|=|\alpha_3|^2+|\alpha_6|^2.
\eeq
The minimum is attained for $\theta=0$ 
(meaning that we can take, without loss of generality, $\alpha_{3},\alpha_{6}\in \mathbb R$). 
The maximum corresponds to $\theta=\pi/2$ (meaning that we can take, without loss of generality, $\alpha_3\in \mathbb R$ and $\alpha_6=-i|\alpha_6|$).
In particular, it follows from Eq. (\ref{boundcoherence}) that a necessary condition for having persistent concurrence is  
\beq\label{condcoherence}
|\alpha_3|\neq |\alpha_6|.
\eeq

\subsection{Asymptotic concurrence}

In order to obtain the amount of fermionic entanglement of the state (\ref{rhoTinft}), we compute the eigenvalues $\Lambda$ of the matrix
 $\rho(t\rightarrow\infty)\mathbb{D} \rho(t\rightarrow\infty) \mathbb{D}^{-1}$, as explained below Eq. (\ref{matrizD}). 
 Direct calculation gives 
\begin{subequations}\label{eigen}
\begin{eqnarray}
\Lambda_1&=&\Lambda_2=\rho_{11}\rho_{55}=|\alpha_1\alpha_5|^2,\\  \Lambda_3&=&\Lambda_4=\rho_{22}\rho_{44}=|\alpha_2\alpha_4|^2,\\
\Lambda_5&=&|\alpha_3|^4 + |\alpha_6|^4 -2\,\textrm{Re}( \alpha_6\alpha_3^{*})^2=\mathcal K^2_{\infty},\\
\Lambda_6&=&0.
\end{eqnarray}
\end{subequations}

Resorting to Eq. \eqref{Cf}, the fermionic entanglement in the long-time regime is $C^{\infty}_f = \max \{ 0, \lambda^{\infty}_1 - \sum_{i=2}^6 \lambda^{\infty}_i\}$,
 where
$\lambda^{\infty}_1$ is the square root of the largest of the eigenvalues  (\ref{eigen}).
If the largest is $\Lambda_1=\Lambda_2$ (so $\lambda^{\infty}_1= \sqrt{\Lambda_1}$), then 
\beq
\lambda^{\infty}_1 - \sum_{i=2}^6 \lambda^{\infty}_i=-2\sqrt{\Lambda_3}-\sqrt{\Lambda_5}\leq 0.
\eeq
The same holds if the largest eigenvalue were $\Lambda_3=\Lambda_4$. 
Consequently,  a necessary condition for having $C^{\infty}_f\neq 0$ is that the largest eigenvalue is $\Lambda_5$.
In such case we get 
\begin{eqnarray}\label{cfTinf1}
  C^{\infty}_f&=& \max \{ 0,\sqrt{\Lambda_5}-2(\sqrt{\Lambda_1}+\sqrt{\Lambda_3})\}\nonumber\\
  &=&\max \{ 0,\mathcal K_{\infty}-2(|\alpha_1\alpha_5|+|\alpha_2\alpha_4|)\},
\end{eqnarray}
so persistent entanglement will exist provided 
\begin{equation}
 \mathcal K_{\infty}>2(|\alpha_1\alpha_5|+|\alpha_2\alpha_4|).
\label{ineq}
\end{equation}
Notice that this implies that for $|\alpha_n\alpha_m|=0$ with $(n,m)=(1,5)$ and $(n,m)=(2,4)$, the concurrence is maximal and equals the amount of coherence.
%
\subsection{Families of states with invariant entanglement}\label{persistent}

With the purpose of identifying families of initial states that evolve with long-lasting concurrence we first focus on the trivial solutions to  (\ref{ineq}), satisfying 
\beq\label{triv}
\alpha_2\alpha_4=\alpha_1\alpha_5=0,\quad \mathcal K_{\infty}\neq 0.
\eeq
The initial states complying with these conditions have the structure
\begin{equation}\label{famII}
     \ket{\psi_0}_{\textrm{I}} =\alpha_r\ket{\psi_r^-}+\alpha_s\ket{\psi_s^-}+\alpha_3\ket{\psi_3^-} + \alpha_6\ket{\psi_6^-},
\end{equation}
where $\alpha_3$ and $\alpha_6$ are subject to the condition (\ref{condcoherence}), and the indices $r,s$ pertain to components of different subspaces of exponential decay, 
i.e., $(r,s)=(1,2),(1,4),(5,2)$, or $(5,4)$. 
For $\alpha_r=\alpha_s=0$ we obtain states in the DF subspace (family DFS), whereas for more general values of these coefficients a family (denoted as I) of states is obtained.
%

We now consider the initial state
\begin{equation} \label{IIex}
    \ket{\psi_0} =\alpha_3\ket{\psi_3^-}+\alpha_4\ket{\psi_4^-} + \alpha_5\ket{\psi_5^-}+\alpha_6\ket{\psi_6^-},
\end{equation}
taken as a representative element of the family I. 
It evolves into a density matrix of the form
\begin{equation}\label{rhoII}
\rho(t)=\left(\begin{array}{cccccc}
0 & 0 & 0 & 0 & 0 &0 \\
0 & 0 &0 & 0& 0 & 0 \\
0 & 0 & \rho_{33} & \rho_{34}(t) & \rho_{35}(t) & \rho_{36} \\
0 & 0 & \rho_{43}(t) & \rho_{44} & \rho_{45}(t) & \rho_{46}(t)\\
0 & 0 & \rho_{53}(t) & \rho_{54}(t) & \rho_{55} & \rho_{56}(t) \\
0 & 0 & \rho_{63} & \rho_{64}(t) & \rho_{65}(t) & \rho_{66} \\
\end{array}\right).
\end{equation}
The matrix $\rho(t)\mathbb{D} \rho(t) \mathbb{D}^{-1}$ has only one non-vanishing and time-independent eigenvalue, given by $\Lambda=\mathcal K^2_{\infty}$.
Consequently, the concurrence of the state (\ref{IIex}) becomes time-independent:
\begin{equation} \label{concufamI}   
C_{f}(t)=C_{f}(0)=\mathcal K_{\infty}=C_{f}^{\infty}.
\end{equation}

\begin{figure}[h]
\includegraphics[width=0.5\columnwidth]{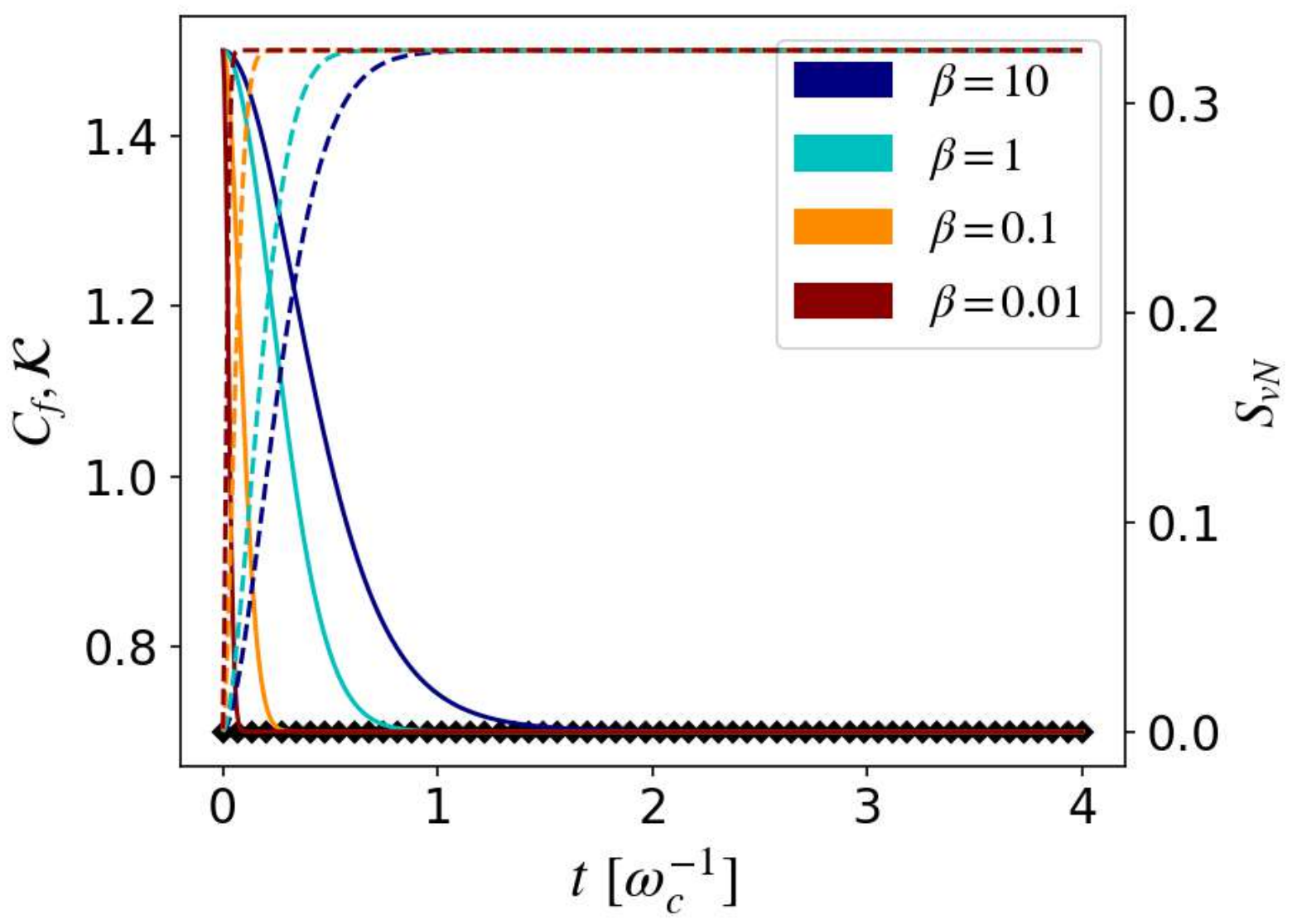}
\includegraphics[width=0.5\columnwidth]{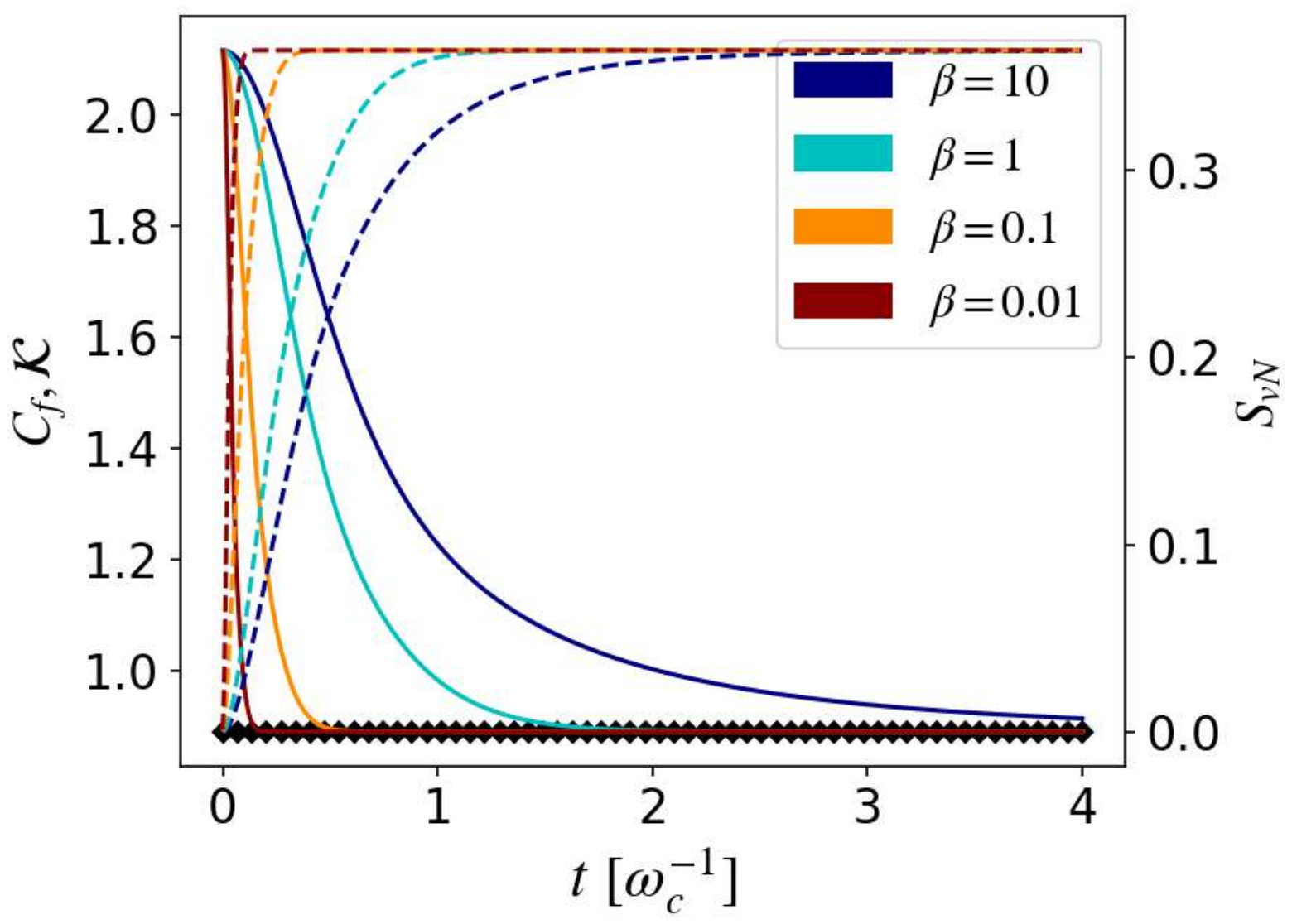}		%
\caption{Concurrence $C_f(t)$ (symbols), coherence $\mathcal K(t)$ (solid lines), and von Neumann entropy $S_{vN}[\rho(t)]$ (dashed lines)  for different temperatures for the initial state \eqref{famII} with:  $\alpha_s=0$, $\alpha_r=\alpha_1=\sqrt{1/10}= \alpha_6$ and $\alpha_3= \sqrt{8/10}$, left panel; 
$\alpha_s=\alpha_4=\sqrt{1/25}$, $\alpha_r=\alpha_5=\sqrt{1/20}$,  $\alpha_6=\sqrt{1/100}$, and $\alpha_3= \sqrt{9/10}$, right panel. Time is measured in units of $\omega^{-1}_c$, and $J_0$ in Eq. (\ref{Gammachange}) was taken as $J_0=8$.}
\label{figfamI}
\end{figure}

An analogous calculation shows that Eq. (\ref{concufamI}) holds for all $\ket{\psi_0}_{\textrm I}$, whence for these states the fermionic entanglement is not only persistent, but invariant. 
This invariance is trivial for states in the family DFS [in line with the discussion below Eq. (\ref{DF})], yet for more general states in the family I, $\rho(t)$ evolves in time and decoherence is present throughout the evolution. 
This is exemplified in Figure \ref{figfamI}, showing the evolution of the entanglement (symbols) and the coherence (solid lines) for two different states in family I, at different temperatures.
We have also plotted  the von Neumann entropy $S_{vN}(\rho)=-\textrm{Tr}\,(\rho \ln\rho)$ of the two-fermion state $\rho(t)$ (dashed-lines). As expected, the dynamics of the latter is opposite to that of the coherence; indeed, as the state decoheres it becomes more mixed, the information of the system is being lost and a concomitant increase in the entropy is observed. In the long-time regime the coherence stabilizes and accordingly the entropy reaches a steady value. 
Since $S_{vN}(\rho)$ is directly related to the information content of the fermionic system as a whole, it  provides no information regarding the (constant) entanglement between the fermions, thus evolves independently of the concurrence
\footnote{It is worth noting that, firstly, if the global (fermionic plus environment)  system were in a pure state, $S_{vN}(\rho)$ would be directly related to the amount of entanglement created between the bath and the two-fermion system. 
Now, since in the case under consideration the global state is not pure but mixed, it would be necessary to resort to entanglement criteria \cite{Horodecki1996,Batle2002} to verify that the fermionic susbystem and the bath become entangled. 
Secondly, as stated in the text, $S_{vN}(\rho)$ alone does not provide information regarding the entanglement between the fermions. The latter can be inferred by comparison of $S_{vN}(\rho)$ with the entropy of the single-fermion state, via appropriate entropic criteria \cite{Zander2012}. However, the use of such criteria is superfluous since we are able to compute the exact amount of fermionic entanglement via the concurrence.}.

\begin{figure}[h]
\includegraphics[width=0.5\columnwidth]{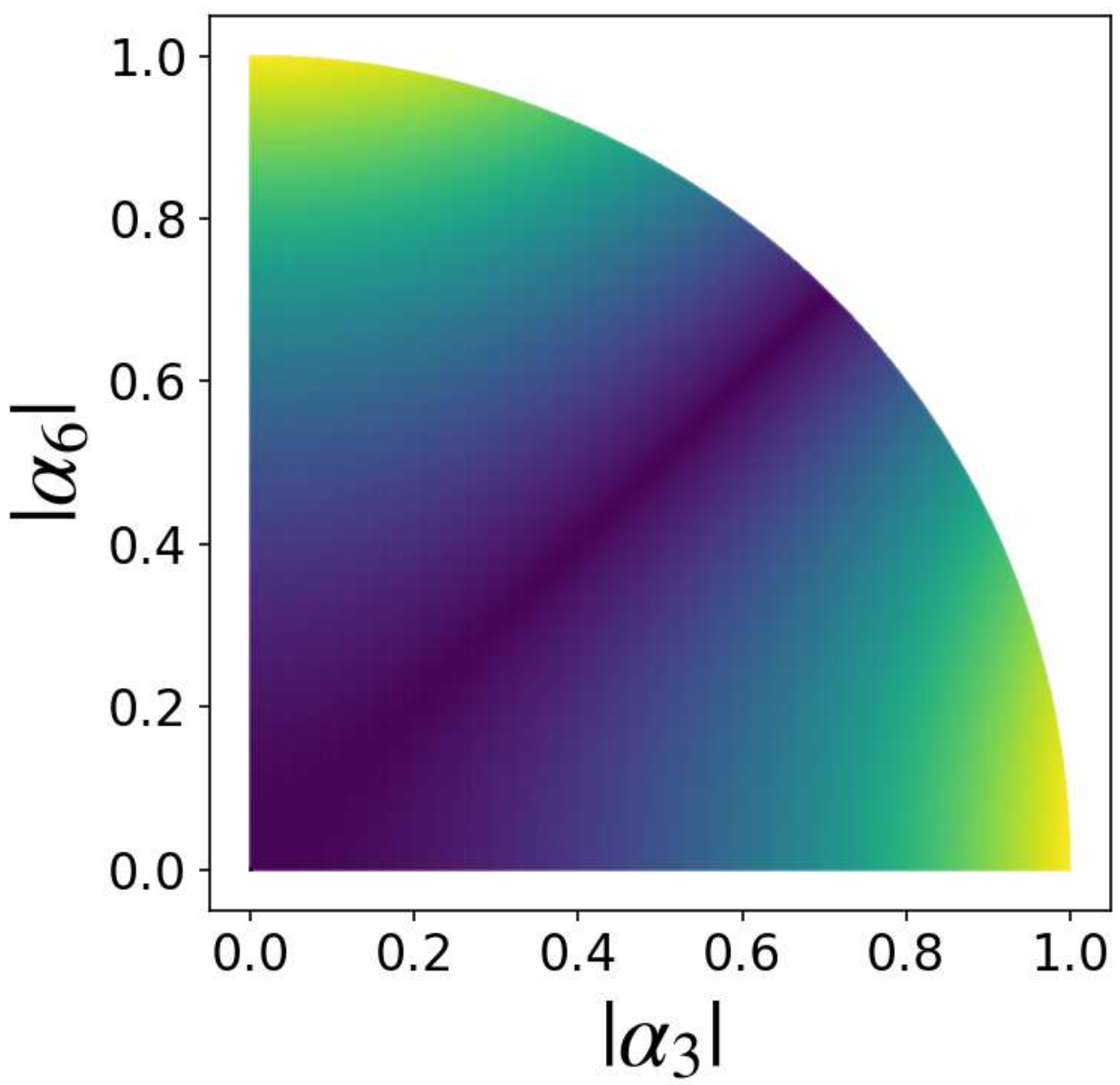}\includegraphics[width=0.5\columnwidth]{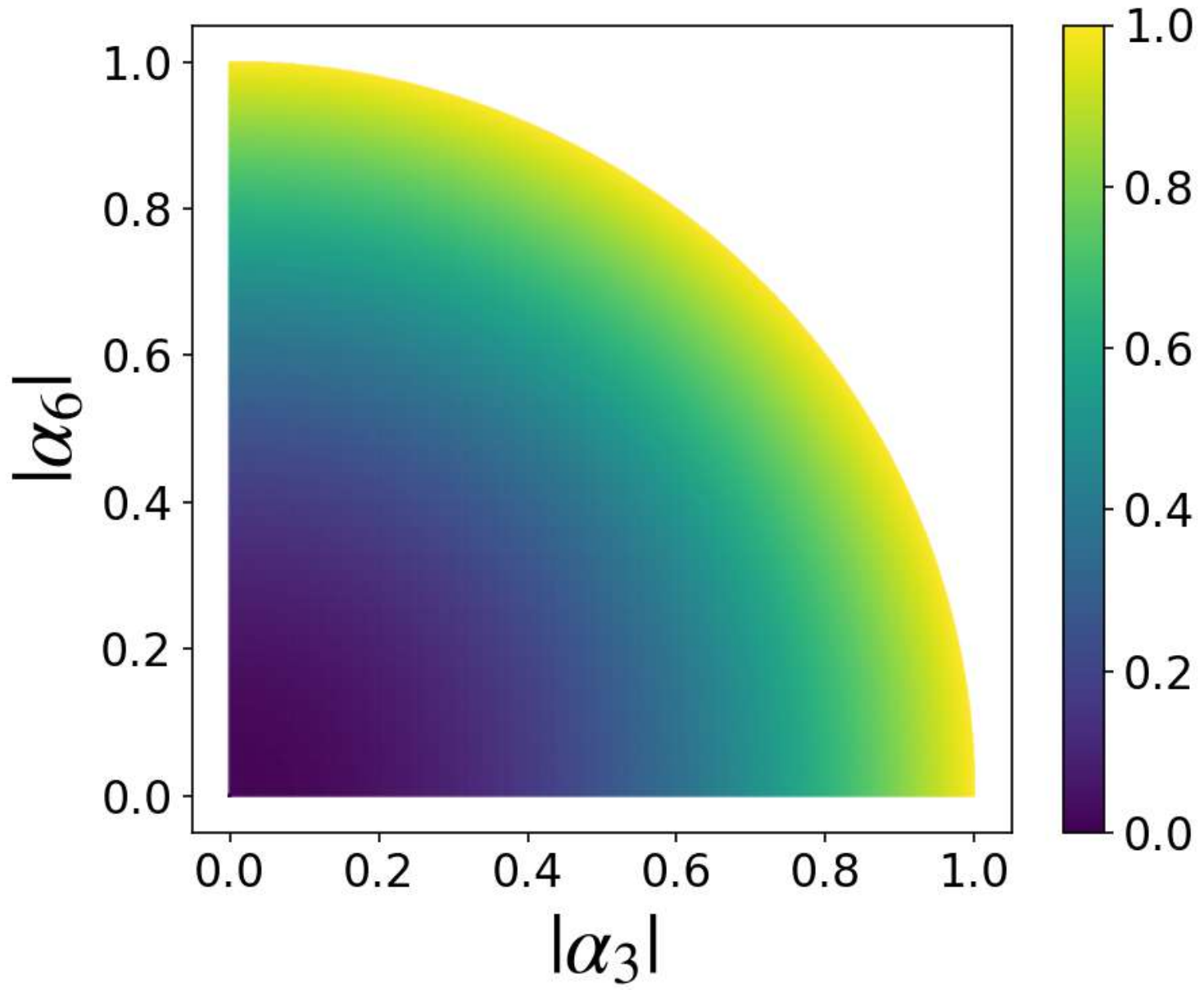}		%
\caption{Fermionic concurrence $C_f=\mathcal K_{\infty}$ for the states $\ket{\psi_0}_{\textrm I}$, Eq. (\ref{famII}). 
The relative phase between $\alpha_3$ and $\alpha_6$ is  $\theta=0$ (left panel, $C_f=\mathcal K^{\min}_{\infty}$) and $\theta=\pi/2$ (right panel, $C_f=\mathcal K^{\max}_{\infty}$).}
\label{fig:enter-label}
\end{figure}

The above results show that the entanglement of the states $\ket{\psi_0}_{\textrm I}$ is fully robust against decoherence. 
Further, the corresponding (invariant) concurrence  (\ref{concufamI}) is the maximal one that can be attained in the long-time regime, as follows from Eq. (\ref{cfTinf1}), and  depends only on $\alpha_3$ and $\alpha_6$, i.e., on the components along the DF subspace.
We can thus characterize $C_f$ for states in the families DFS and I, as shown in Figure \ref{fig:enter-label}, which depicts $C_f$ in color scale as a function of  $|\alpha_3|$ and $|\alpha_6|$. 
The left panel illustrates $C_f$ assuming positive  values for these coefficients ($\theta=0$), which corresponds to $C_f=\mathcal K^{\min}_{\infty}=|\alpha_3^2-\alpha_6^2|$, that is, to the minimum concurrence as follows from Eqs. (\ref{boundcoherence}) and (\ref{concufamI}).
This plot clearly shows signatures of a single Slater determinant (vanishing fermionic entanglement) along the diagonal $\alpha_3=\alpha_6=\alpha$. 
In this case the states in the family DFS become the 
equally weighted superposition $\tfrac{1}{\sqrt{2}}(|\psi_3^-\rangle+|\psi_6^-\rangle)=|\psi^{sl}_{14}\rangle$. 
As for the states (\ref{famII}) of the family I, it is straightforward to prove that when the states $|\psi^-_r\rangle$ and $|\psi^-_s\rangle$
are added to the vector $\alpha(|\psi_3^-\rangle+|\psi_6^-\rangle)\sim|\psi^{sl}_{14}\rangle$, a superposition of non-orthogonal Slater determinants is obtained, which can be written as a single Slater determinant, corresponding to a separable state. 
%
In contrast with the diagonal line, as $|\alpha_3|\rightarrow 1$, hence $|\alpha_6|\rightarrow 0$ (or vice versa), we get the maximally entangled state $|\psi^{-}_{3}\rangle$ (or $|\psi^{-}_{6}\rangle$).

The right panel in Fig. \ref{fig:enter-label} displays $C_f=\mathcal K^{\max}_{\infty}=|\alpha_3|^2+|\alpha_6|^2$, that is, the maximum concurrence as follows from Eqs. (\ref{boundcoherence}) and (\ref{concufamI}),
corresponding to coefficients \(\alpha_3\) and \(\alpha_6\) with a relative phase \(\theta = \pi/2\) 
(in particular we chose $\alpha_3\in \mathbb R$ and \(\alpha_6 = i|\alpha_6|\)). 
The plot exhibits a clear radial symmetry, with the radius corresponding precisely to the value of $C_f$, so maximally entangled states ($C_f=1$) lie along the maximal circumference.

\subsection{Persistent concurrence}

We now perform a numerical sampling of the concurrence \eqref{cfTinf1} considering more general states $\ket{\psi_0}$ (other than those pertaining to the families DFS and I). 
From now on we will assume real coefficients $\alpha_n\geq 0$. 
Notice that taking $\alpha_3,\alpha_6\in \mathbb R$ amounts to consider the most conservative scenario, corresponding to the minimum coherence given by $\mathcal K^{\min}_{\infty}=|\alpha_3^2-\alpha_6^2|$, as follows from Eq. (\ref{boundcoherence}).
For simplicity in the writing, we introduce the notation
\begin{subequations}
\begin{eqnarray}\label{xyz1}
    x&=& |\alpha_2\alpha_4|=\alpha_2\alpha_4, \\ 
    y&=& |\alpha_1\alpha_5|=\alpha_1\alpha_5, \\ 
    z&=& |\alpha_3^2-\alpha_6^2|=\mathcal K^{\min}_{\infty},
\end{eqnarray}
\end{subequations}
in terms of which the concurrence in the long-time regime writes as
\begin{equation}\label{Cxyz}
    C^{\infty}_f = \max \{0, z-2(x+y)\}.
\end{equation}
In this way any initial state, characterized by the set of coefficients $\{\alpha_n\}$, will be mapped into a point $\boldsymbol{r}=(x,y,z)$ 
The map $\ket{\psi_0}\leftrightarrow \boldsymbol{r}$ is not one-to-one, since infinitely many initial states will correspond to the same $\boldsymbol{r}$.
In the present case with $\alpha_n\geq 0$, 
the initial concurrence of all such states that are mapped into the same $\boldsymbol{r}$ admits only two different values, depending on whether $\alpha_3\geq\alpha_6$ or $\alpha_3<\alpha_6$ 
[for states with $\alpha_3\geq\alpha_6$ we get $z=\alpha_3^2-\alpha^2_6$, and 
 Eq. (\ref{concurrencef}) gives 
 $C_f(0)=|z+2(y-x)|$, whereas for states with $\alpha_3<\alpha_6$, $z=\alpha_6^2-\alpha^2_3$ and $C_f(0)=|-z+2(y-x)|$].
 %
In the forthcoming analysis we will assume (without loss of generality) that $\alpha_3 \geq\alpha_6$. 
%

\begin{figure}[h!]
\includegraphics[width=0.55\columnwidth]{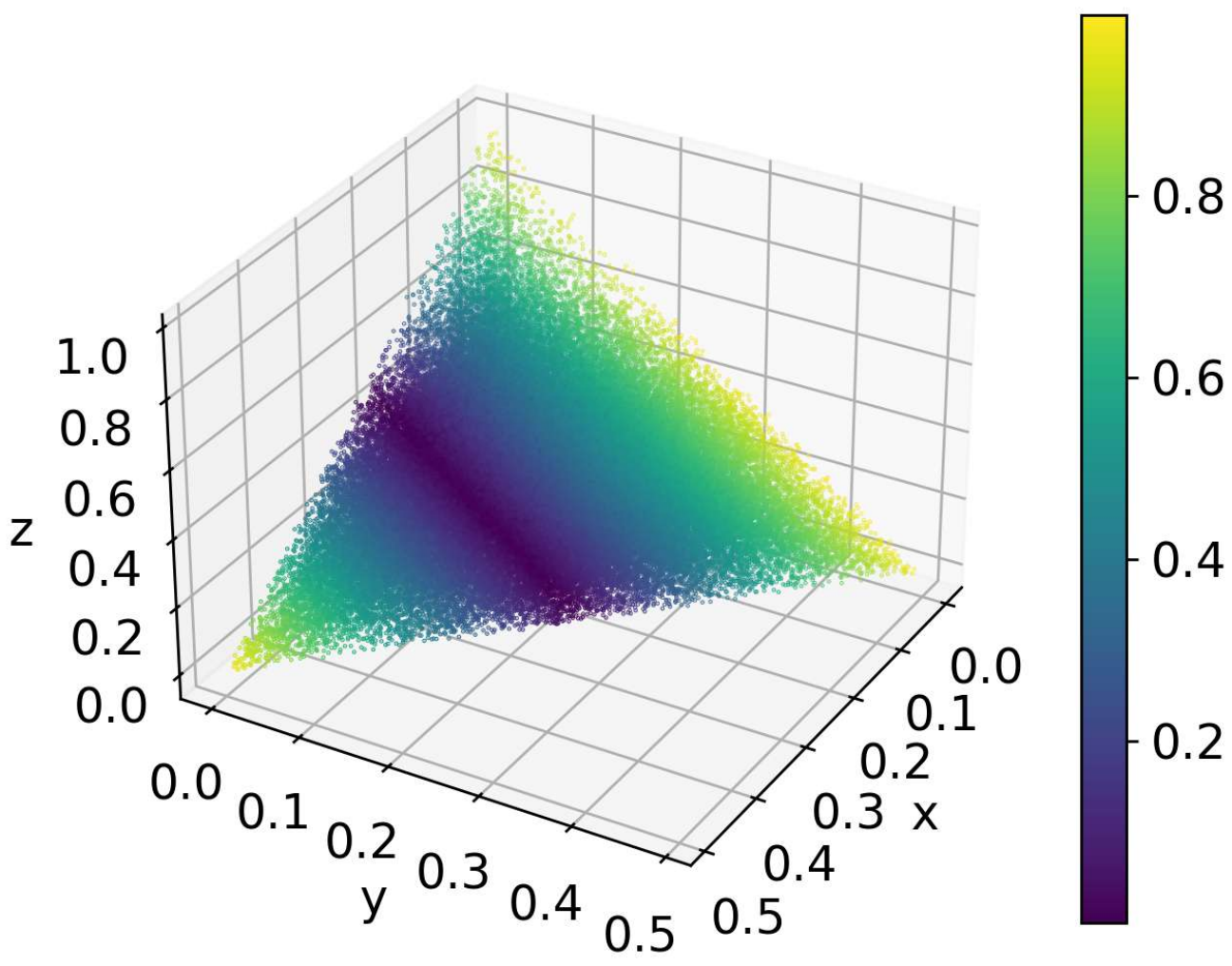}
\includegraphics[width=0.55\columnwidth]{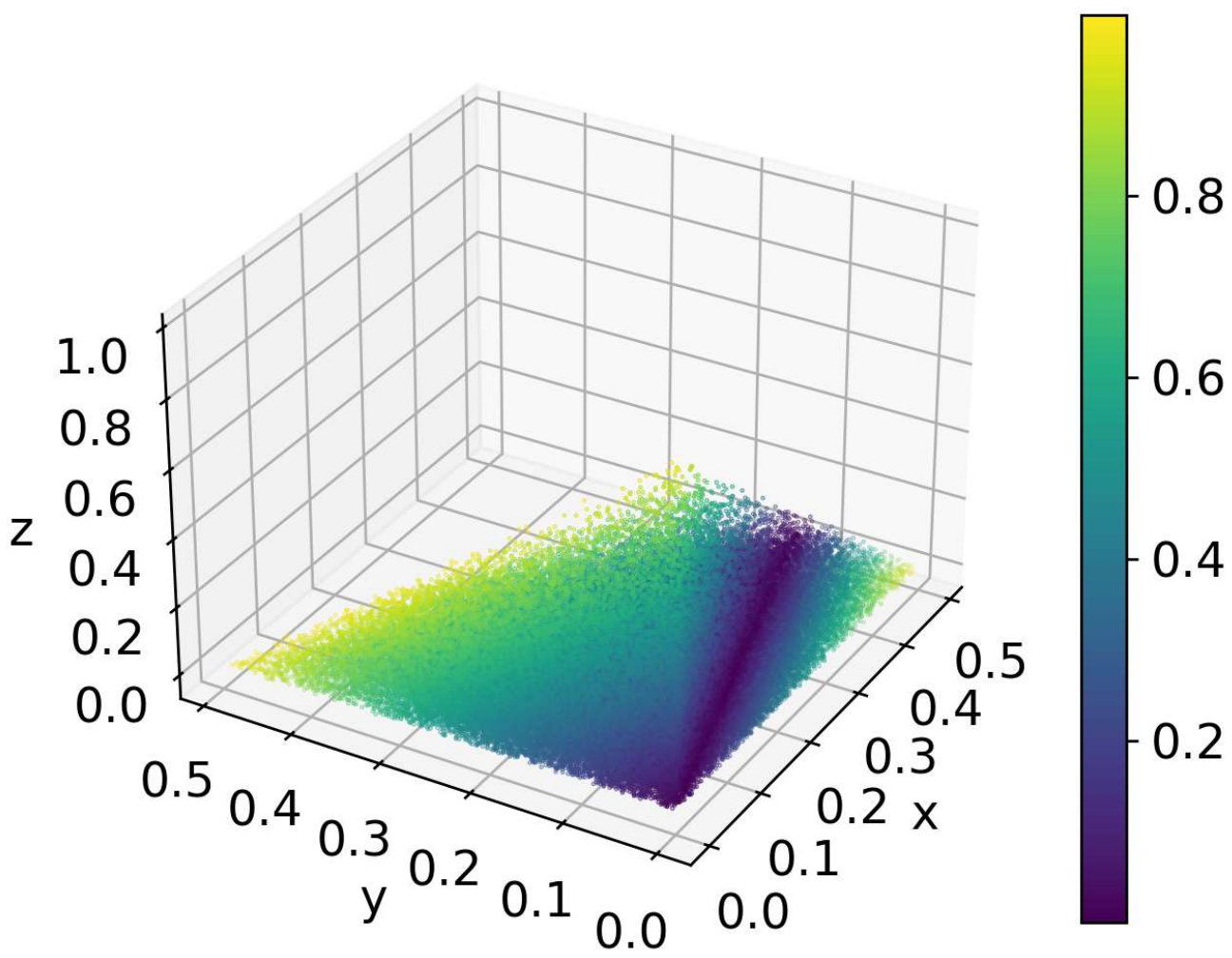} 
  \caption{Initial concurrence 
$C_f(0)$ (color scale) in the space $(x,y,z)$ as seen from two different perspectives.}
\label{piramideinicial}
\end{figure}

\begin{figure}[h]
\includegraphics[width=0.5\columnwidth]{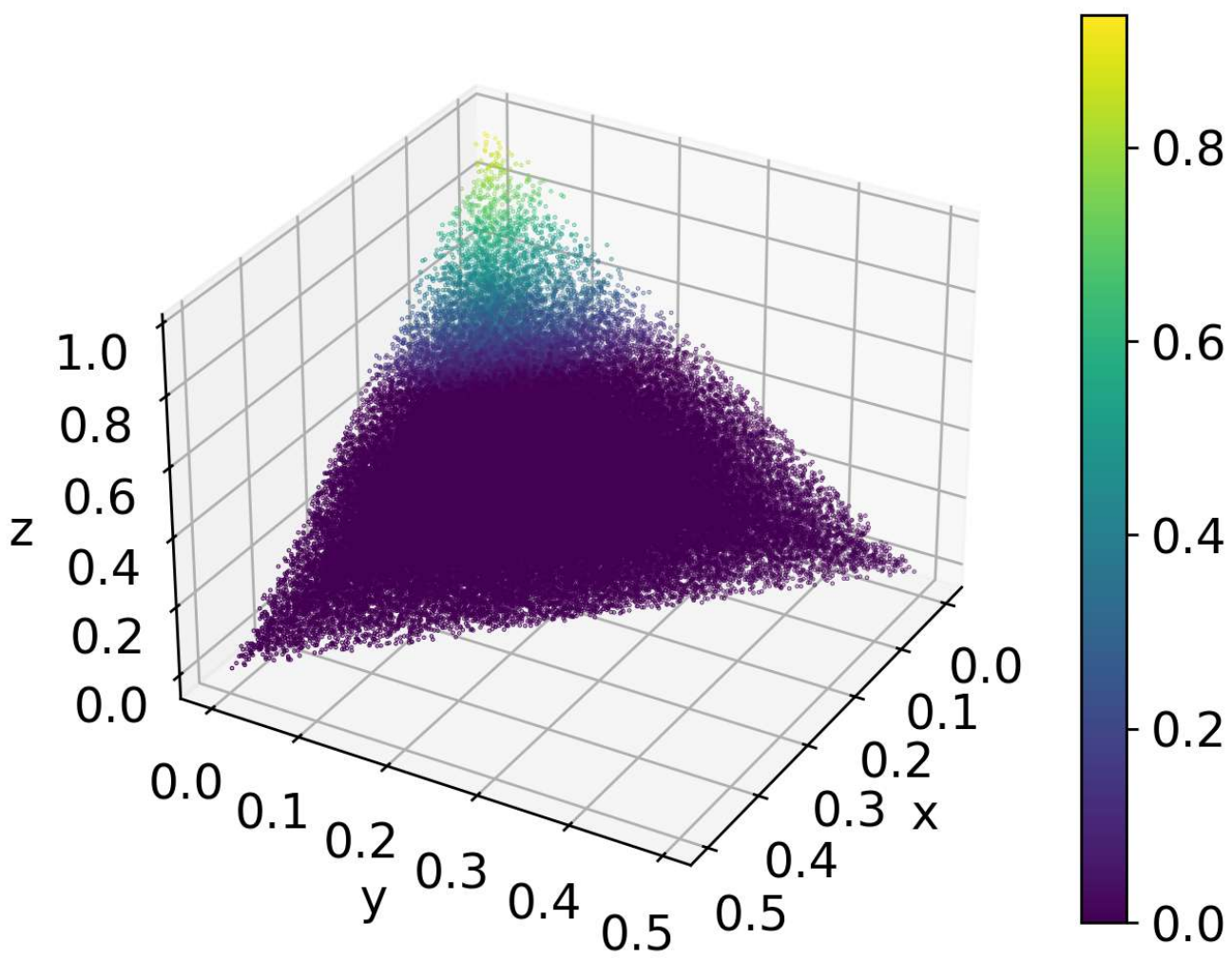}
\includegraphics[width=0.5\columnwidth]{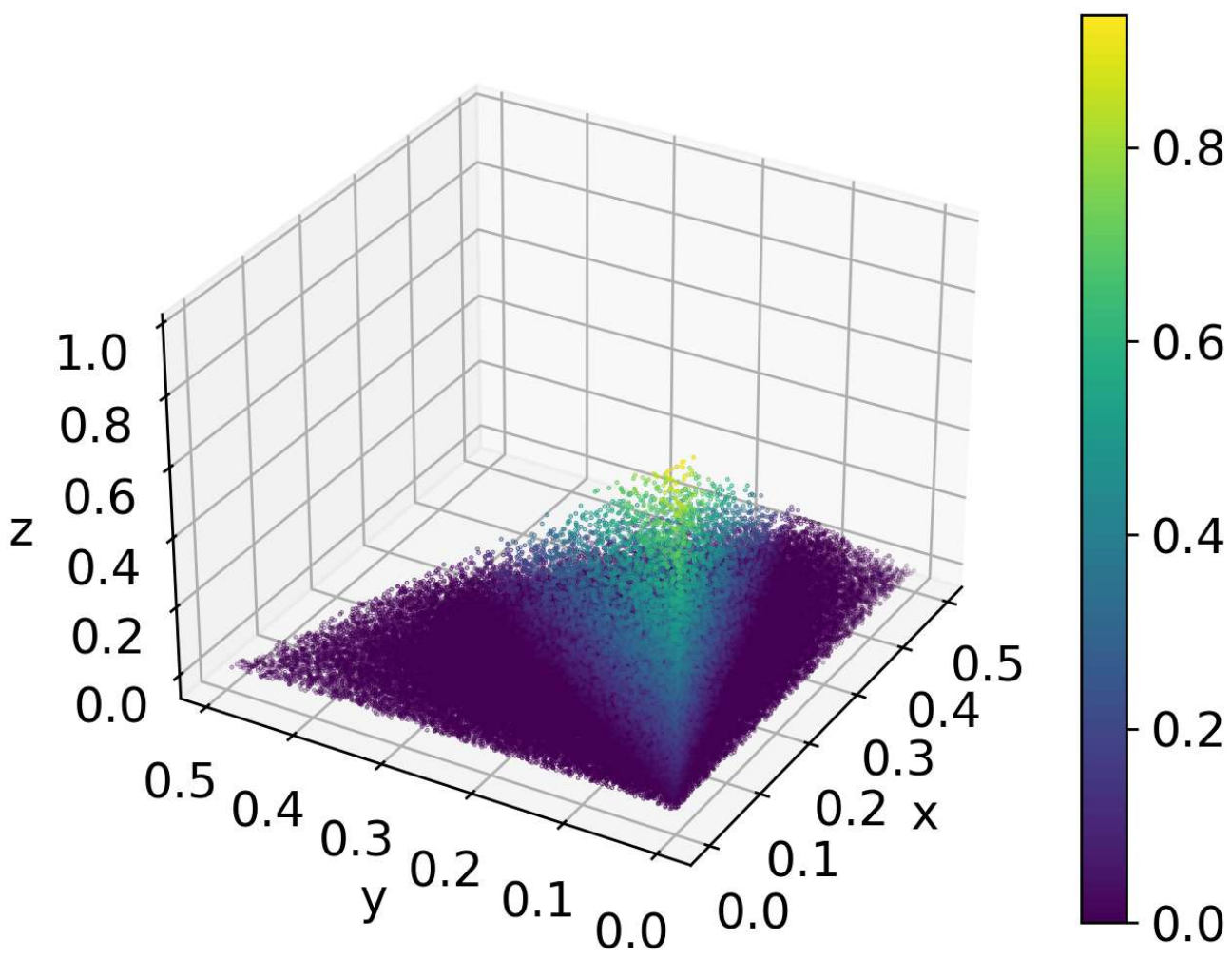} 
  \caption{Asymptotic concurrence 
$C^{\infty}_f$ (color scale) in the space $(x,y,z)$ as seen from two different perspectives.}
\label{piramidefinal}
\end{figure}

\begin{figure}[h!]
\includegraphics[width=0.5\columnwidth]{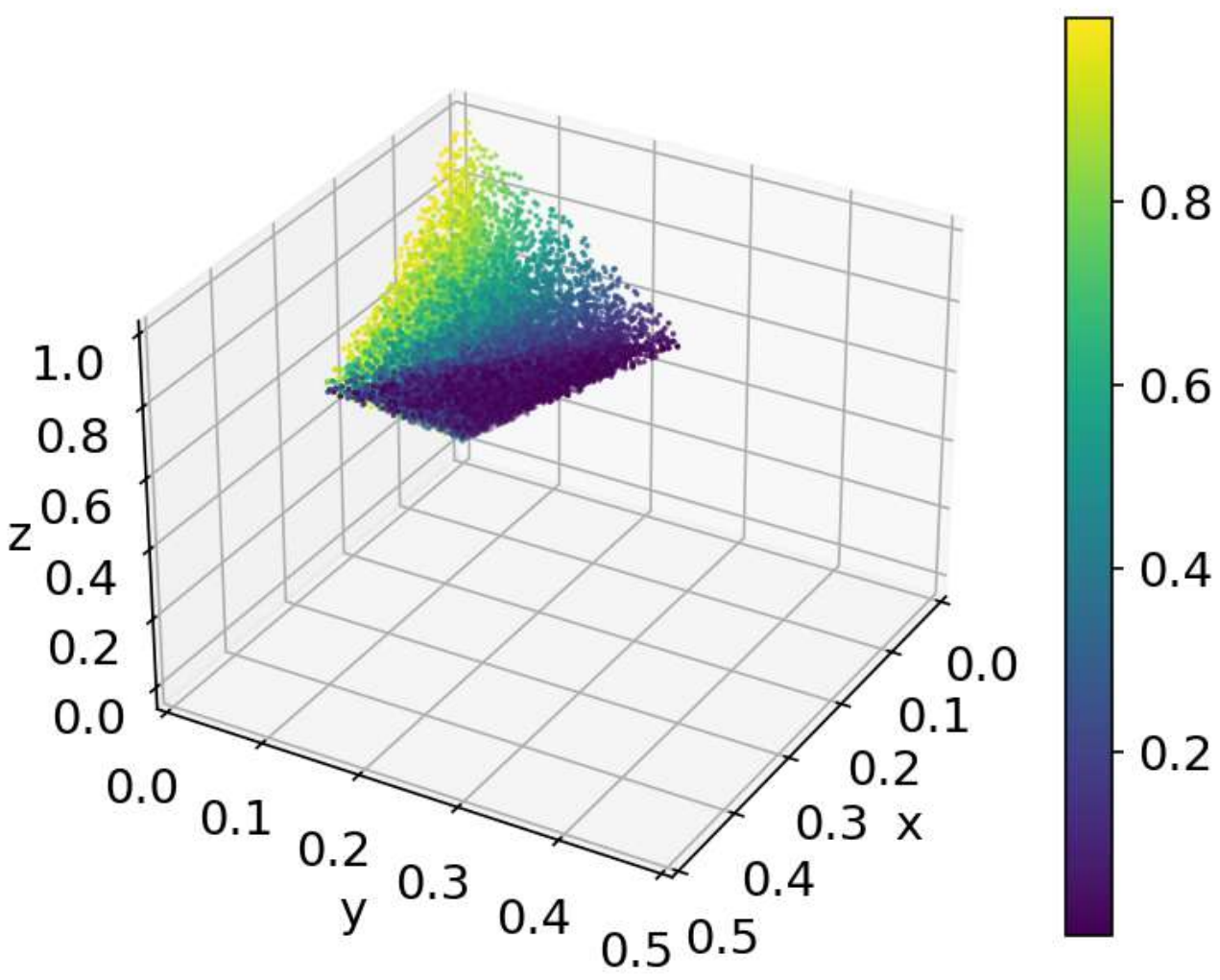} %
\includegraphics[width=0.5\columnwidth]{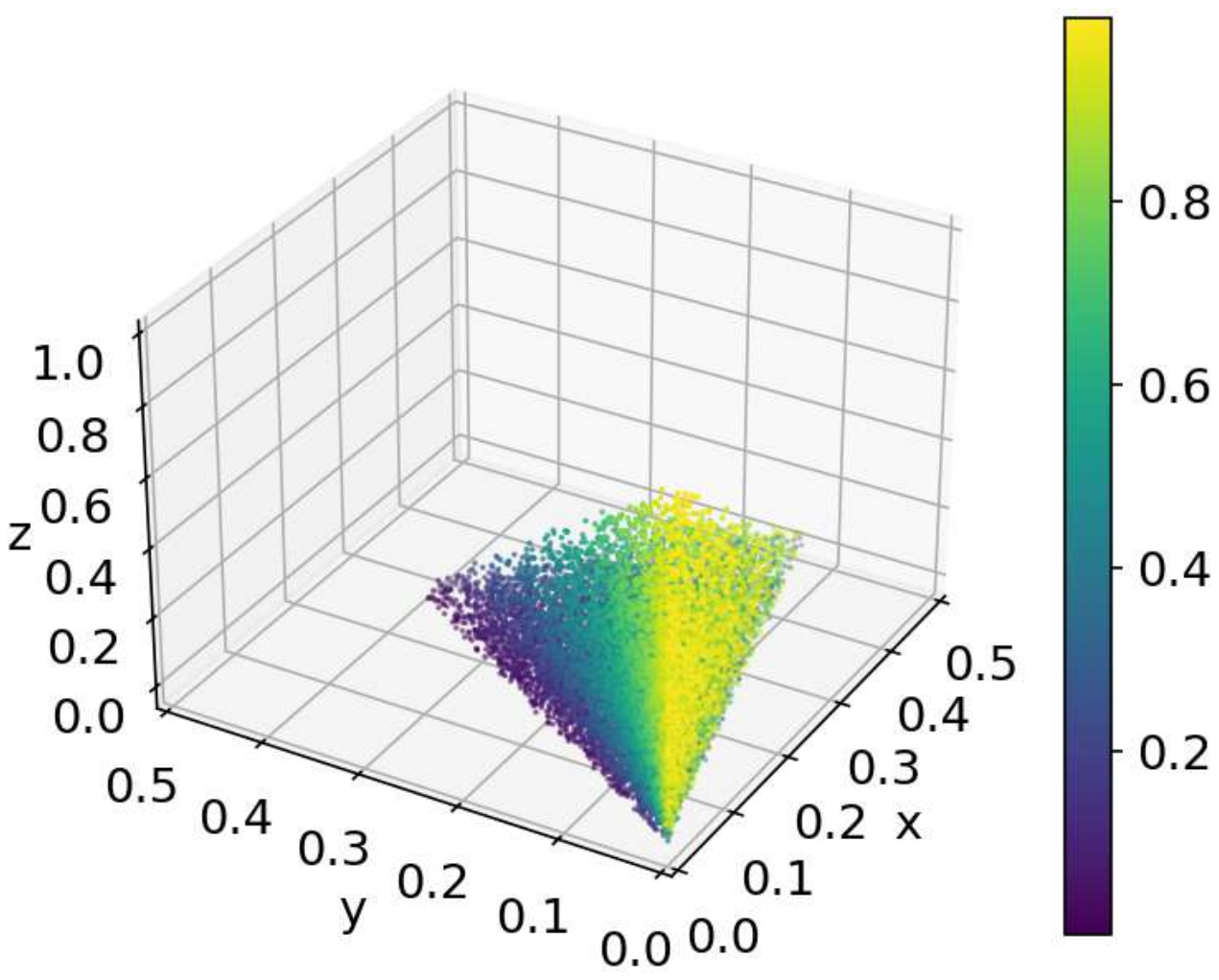} 
\caption{Persistence $P$ (color scale) in the space $(x,y,z)$ as seen from two different perspectives. Only points with non-zero persistence are shown. }
\label{figP}
\end{figure}

In order to compare the initial and final entanglement between the fermions, we: 
1) Sample $1\times 10^5$ initial states ($1\times 10^5$ sets $\{\alpha_n\}$ complying with $\sum_n|\alpha_n|^2=1$) with $\alpha_3 \geq\alpha_6$; 
2) Compute the initial concurrence from Eq. (\ref{concurrencef}); 
3) Determine the coordinates $x,y,z$ via Eqs. (\ref{xyz1}), and the asymptotic concurrence using Eq. (\ref{Cxyz});
4) Plot in color scale, and at each point $(x,y,z)$, the initial concurrence $C_f(0)$ (Fig. \ref {piramideinicial}), the asymptotic concurrence $C^{\infty}_f$ (Fig. \ref {piramidefinal}), and the ratio $C^{\infty}_f/C_f(0)$ (Fig. \ref {figP}).

Figures \ref{piramideinicial} and \ref{piramidefinal} show that all the points $\boldsymbol{r}$ lie within a tetrahedron 
determined by $0\leq 2(x+y)+z\leq 1$ (consistently with the normalization condition $\sum_n|\alpha_n|^2\leq 1$).
The vertices are located at:  $(0,0,0)$, corresponding to $C_f(0)=C^{\infty}_f=0$;  $(\tfrac{1}{2},0,0)$ and $(0,\tfrac{1}{2},0)$, both corresponding to $C_f(0)=1$ and $C^{\infty}_f=0$; 
$(0,0,1)$
 corresponding to $C_f(0)=C^{\infty}_f=1$.

As seen from Fig. \ref{piramideinicial}, the hypotenuse of the tetrahedron's face located at $x=0$, determined by the line $z=1-2y$,  corresponds to $C_f(0)=1$, in line with Eq. (\ref{concurrencef}). 
%
The states $\ket{\psi_0}_{\textrm I}$ are mapped into points in the $z$ axis, and since these posses invariant entanglement it holds that $C_f(0)=C_f(t)=z$.  
States (\ref{expdecay}) in the Exponential Decay subspace are mapped into the $x$ axis for $(n,m)=(2,4)$, and into the $y$ axis for $(n,m)=(1,5)$. The corresponding initial concurrence is, respectively, $C_f(0)=2x$ and $C_f(0)=2y$.  

Figure \ref{piramidefinal} shows the concurrence in the long-time regime.
As expected, it exhibits a symmetry with respect to the plane $x=y$ (since $C_f^{\infty}$ is invariant under the exchange $x\leftrightarrow y$, as seen from Eq. (\ref{Cxyz})). 
We also observe that the asymptotic concurrence increases as we approach the vertex $(0,0,1)$, corresponding to the maximally entangled state $\ket{\psi_3^-}$.
%
%
As seen above, the points along the $z$ axis  correspond to states with invariant entanglement, whence the color scale in such axis in Figs. \ref{piramideinicial} and \ref{piramidefinal}  is identical.
As expected, the points lying along the axes $x$ and $y$, corresponding to the ED subspace, have null entanglement in the long-time regime. 

The hypotenuse of the tetrahedron's face located at $x=0$, determined by the line $z=1-2y$ and corresponding (as seen above) to maximal initial entanglement, exhibits a gradual decay in the fermionic concurrence in the asymptotic regime, given by $C^{\infty}_f=1-4y$. 
It is maximal at the point $(0,0,1)$ and vanishes at $(0,\tfrac{1}{4},\tfrac{1}{2})$; 
from that point on, the entanglement is null all along the segment $(0,y,1-2y)$ for $y\in[\tfrac{1}{4},\tfrac{1}{2}]$. 
The same goes for the hypotenuse of the tetrahedron's face located at $y=0$, determined by the line $z=1-2x$.

Figure \ref{piramidefinal} shows that a large volume of the tetrahedron corresponds to states that lose all their initial entanglement, yet a considerable fraction of points exists that reveal the presence of states with persistent concurrence.
In line with Eq. (\ref{Cxyz}), these states are mapped into points within the original tetrahedron that comply in addition with $z>2(x+y)$.
To quantify the persistence of entanglement we focus on the ratio
\begin{equation}
    P=\frac{C_f(\infty)}{C_f(0)},\quad(C_f(0)\neq 0),
\end{equation}
which vanishes whenever the asymptotic concurrence is null (no entanglement persisted), and is maximal ($P=1$) when no entanglement was lost. 
In order to verify that $P \leq 1$ we first resort to the reverse triangle inequality in Eq. (\ref{concurrencef}) to get
\begin{eqnarray}\label{conc0}
   C_f(0) &\geq& \Big||\alpha_3^2-\alpha_6^2|- 2|\alpha_1\alpha_5-\alpha_2\alpha_4|\Big|\nonumber\\
   &\geq&|\alpha_3^2-\alpha_6^2|- 2|\alpha_1\alpha_5-\alpha_2\alpha_4|\nonumber\\
   &\geq&|\alpha_3^2-\alpha_6^2| -2 (|\alpha_1\alpha_5|+|\alpha_2\alpha_4|),
\end{eqnarray}
where in the last line we used the triangle inequality.
Now, the first term reads $|\alpha_3^2-\alpha_6^2|=\sqrt{|\alpha_3|^4+|\alpha_6|^4-2|\alpha_3|^2|\alpha_6|^2\cos2\theta}=\mathcal K_{\infty}$,
so using Eq. (\ref{cfTinf1}) we are led to
\begin{equation}
C_f(0)\geq      \mathcal K_\infty -2 (|\alpha_1\alpha_5|+|\alpha_2\alpha_4|) \geq C_f(\infty).
\end{equation}
This shows that, as expected, the fermionic concurrence in the long-time regime cannot exceed its initial value, and therefore $P\leq 1$ (note that the conclusion is general enough and does not require the assumption of real coefficients). 
%

Figure \ref{figP} shows the region with $P>0$.
It is observed that the face at $y=0$ has $P=1$, so all its points correspond to states whose entanglement endures in the long-time regime.
For $y=0$, Eqs. (\ref{concurrencef}) and (\ref{Cxyz}) give $C_f(0)=\max\{0,z-2x\}$ and $C_f(\infty)=|z-2x|$ (where we used that in the present case $z=\alpha^2_3-\alpha^2_6$). Thus, for $y=0$ and $z>2x$ the persistence $P$ is maximal, as seen in the yellow face in the bottom panel of Fig. \ref{figP}.
%
%
\begin{figure}[h!]
\includegraphics[width=0.75\columnwidth]{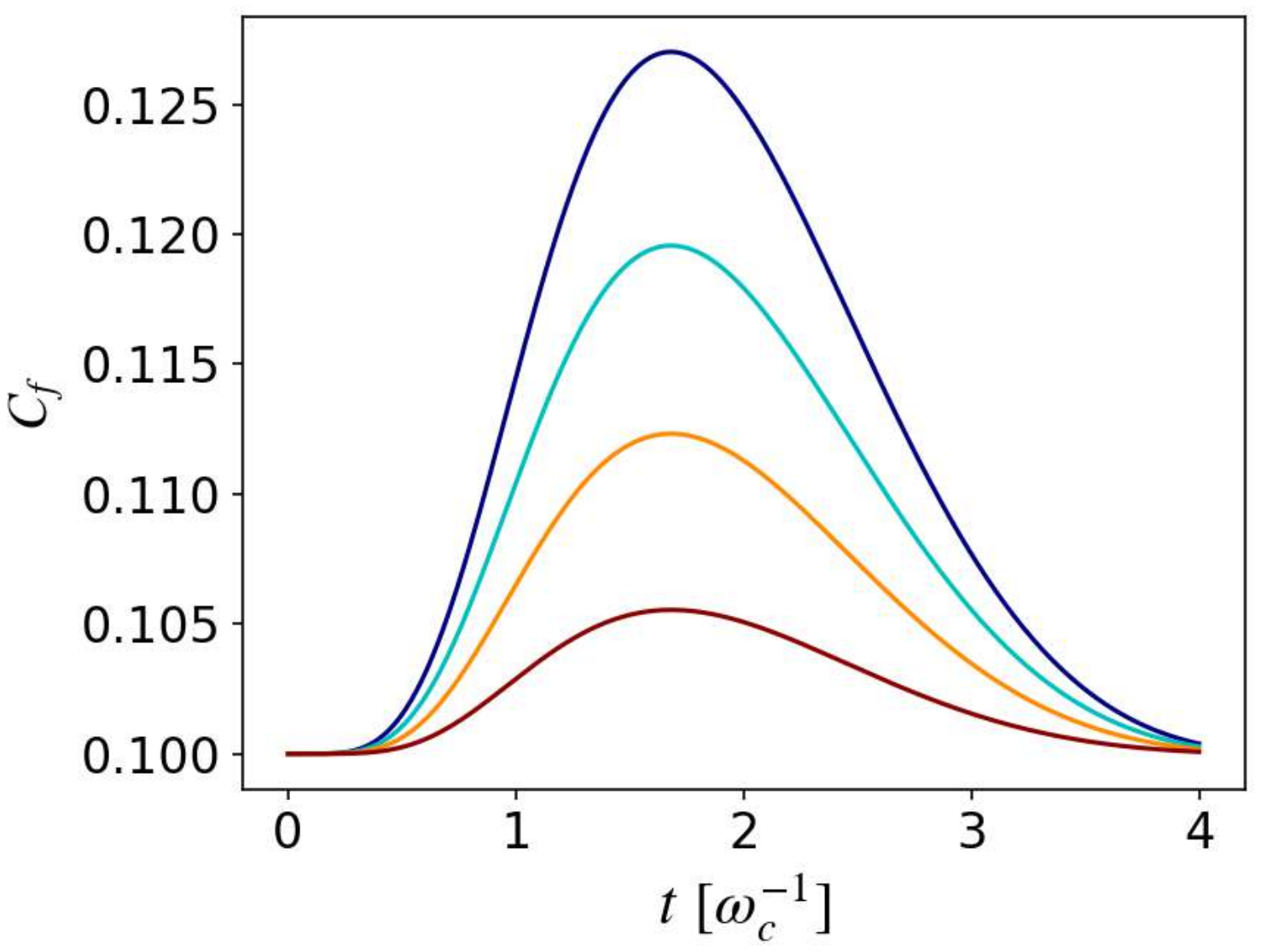}
 \caption{Persistent (though not invariant) concurrence for four states (see text) with $y=0$ at fixed temperature corresponding to $\beta=10$.
 In all cases $C_f(0)=C_f^\infty=1/10$. Time is measured in units of the inverse cutoff frequency $\omega_c$, and $J_0$ in Eq. (\ref{Gammachange}) was taken as $J_0=8$.}
 \label{f8}
\end{figure}

It should be stressed, however, that total persistence ($P=1$) does not necessarily imply invariant concurrence. 
As an example, Fig. \ref{f8} shows the evolution of $C_f$ at fixed temperature ($\beta=10$), for four initial states with $y=0$, hence satisfying $C_f(0)=C_f(\infty)$. 
The states have the structure $\ket{\psi_0}=\alpha_1\ket{\psi_1^-}+\alpha_2\ket{\psi_2^-}+\alpha_3\ket{\psi_3^-}+\alpha_4\ket{\psi_4^-}$ with $\alpha_2=\alpha_4$, and were chosen with the set  $\{\alpha_1,\alpha_2,\alpha_3\}$ equal to: $\Big\{\frac{1}{\sqrt{10}},\frac{1}{\sqrt{5}},\frac{1}{\sqrt{2}}\Big\}$ blue curve;
 $\Big\{\sqrt{\frac{3}{10}},\sqrt{\frac{3}{20}},\sqrt{\frac{2}{5}}\Big\}$ turquoise curve;
 $\Big\{\frac{1}{\sqrt 2},\frac{1}{\sqrt 10},\sqrt{\frac{3}{10}}\Big\}$ orange curve;
 $\Big\{\sqrt{\frac{7}{10}},\frac{1}{\sqrt 20},\frac{1}{\sqrt 5}\Big\}$ red curve. 

\subsection{Persistent entanglement in other decoherence channels}

So far, we showed that under the collective non-dissipative dynamical model governed by the Hamiltonian (\ref{Hamil}) subspaces with long-lived entanglement exist. 
Then the question arises whether entanglement can also persist under other types of decoherence processes.
To advance into this question, first it should be stressed that the previous identification of states with robust entanglement relies solely on the structure of the fermionic state  (\ref{rhoTinft}), so any dynamical evolution leading to a similar density matrix in the long-time regime allows in principle for the existence of long-lasting entanglement.
Of course, other structures of $\rho(t\rightarrow\infty)$ may also be consistent with the existence of states with persistent concurrence.
To exemplify this, we will now briefly consider a paradigmatic \emph{dissipative} decoherence process (so $[H_S,H]\neq 0$), corresponding to the so-called Amplitude Damping Channel (ADC). 

The (standard) ADC represents the dissipative interaction between a qubit and its environment, and is particularly relevant for  modeling the spontaneous atomic decay \cite{Aolita2015}. The dynamics of a pair of qubits each subject to an ADC has been extensively studied, exhibiting fully decoherence, or in some cases entanglement sudden death  \cite{Yu2009,Salles2008,Aolita2015}.
Based on the characteristic features of the ADC acting on a qubit, the corresponding quantum map has been  generalized in \cite{Valdes2015} to the 6-dimensional joint Hilbert space that describes the composite $S$ of two 4-level fermions, as those considered here. 
Thus, assuming that in the two-fermion states $|2,m\rangle$ [see Eqs. (\ref{tablaedos})], $m$ stands for an excitation that can be exchanged with the environment $E$, the AD map reads \cite{Valdes2015}
\begin{subequations}\label{ADchannel}
\begin{eqnarray}
|0,0\rangle_{S}|0\rangle_E&\rightarrow&|0,0\rangle_{S}|0\rangle_E,\\
|2,-2\rangle_{S}|0\rangle|_E&\rightarrow&|2,-2\rangle_{S}|0\rangle_E,\nonumber
\end{eqnarray}
and
\beq
|2,m\rangle_{S}|0\rangle_E\rightarrow\sqrt{1-p}|2,m\rangle_{S}|0\rangle_E+\sqrt{p}|2,m-1\rangle_{S}|1\rangle_E,
\eeq
\end{subequations}
for $m=-1,...,2$, 
where $p=p(t)\in [0,1]$ is a continuous parametrization of time, such that $p(0)=0$. In a decay model, which will be the one considered here, $p$ decays exponentially in time, so $p(\infty)=1$ \cite{Aolita2015}. 

For an initial state $\ket{\Psi(0)}=\ket{\psi_0}_S\otimes\ket{0}_E$, with $\ket{\psi_0}$ given by the generic two-fermion state (\ref{psi_init}), the map (\ref{ADchannel}) gives the evolved state
\beq\label{evolvedADC}
\ket{\Psi(p)}=\ket{A(p)}\ket{0}_E+\ket{B(p)}\ket{1}_E,
\eeq
with
\begin{subequations}\label{ApBp}
  \begin{align}
\ket{A(p)}&= \sqrt{1-p}\Big(\sum^4_{n=1} \alpha_n\ket{\psi^-_n}\Big)+\alpha_5 \ket{\psi^-_5}+  \alpha_6 \ket{\psi^-_6},\label{Ap}\\
\ket{B(p)}&= \sqrt{p}\Big(\sum^4_{n=1} \alpha_n\ket{\psi^-_{n+1}}\Big).
\end{align}  
\end{subequations}
Taking the partial trace over the degrees of freedom of $E$ on $\ket{\Psi}\!\bra{\Psi}$ results in the two-fermion density operator $\rho(p)=\ket{A(p)}\!\bra{A(p)}+\ket{B(p)}\!\bra{B(p)}$.
In the long-time regime (attained as $p\rightarrow 1$), the corresponding density matrix in the basis $\{\ket{\psi^-_n}\}$ reads 
\begin{equation}\label{rhoTinftADC}
\rho(p\rightarrow 1)=\left(\begin{array}{cccccc}
0 & 0 & 0 & 0 & 0 & 0 \\
0 & |\alpha_1|^2 & \alpha_1\alpha_2^* & \alpha_1\alpha_3^* & \alpha_1\alpha_4^* & 0 \\
0 & \alpha_2\alpha_1^* & |\alpha_2|^2 & \alpha_2\alpha^*_3 & \alpha_2\alpha^*_4 & 0 \\
0 & \alpha_3\alpha^*_1 & \alpha_3\alpha^*_2 & |\alpha_3|^2 & \alpha_3\alpha^*_4 & 0 \\
0 & \alpha_4\alpha^*_1 & \alpha_4\alpha^*_2 & \alpha_4\alpha^*_3 & |\alpha_4|^2+|\alpha_5|^2 & \alpha_5\alpha_6^* \\
0 & 0 & 0 & 0 & \alpha_6\alpha_5^* & |\alpha_6|^2
\end{array}\right).
\end{equation}

The long-term effect of the $S$-$E$ interaction on the states populations is apparent from Eq. (\ref{rhoTinftADC}): the population of the state $\ket{\psi^-_i}$ vanishes for $i=1$, varies in general for $i=2,3,4$, does not decrease for $i=5$, and remains constant for $i=6$. 
We also observe that only the coherence term involving the states $\ket{\psi_5^-}$ and $\ket{\psi_6^-}$ coincide with its initial value. 
This was foreseeable from Eq. (\ref{Ap}), indicating that under the amplitude damping dynamics
the vectors $\ket{\psi_0}^{(\textrm{ADC})}_\textrm{DF}= \alpha_5\ket{\psi_5^-} + \alpha_6\ket{\psi_6^-}$ conform a decoherence-free subspace   whose elements are immune to the interaction with the environment, in complete analogy with the states in Eq. (\ref{DF}).  
Yet, unlike what occurred in the previously considered dynamics, in this case the DF subspace is not the only one that contributes to maintain coherence in the long-time regime.
Under the amplitude damping new subspaces of persistent coherence emerge, as can be inferred from the expression of $\rho(p\rightarrow1)$ in the Slater basis, namely 
\begin{equation}\label{rhoTinftADC sl}
\rho^{sl}(p\rightarrow 1)=
\left(
\begin{array}{cccccc}
 0 & 0 & 0 & 0 & 0 & 0 \\
 0 & |\alpha_1|^2 & \frac{1}{\sqrt{2}}\alpha_1 \alpha_2^* & \alpha_1 \alpha_3^* & \alpha_1 \alpha_4^* & \frac{1}{\sqrt{2}}\alpha_1 \alpha_2^* \\
 0 & \frac{1}{\sqrt{2}}\alpha_1^* \alpha_2 & \frac{1}{2}(|\alpha_2|^2+|\alpha_6|^2) & \frac{1}{\sqrt{2}}\alpha_2 \alpha_3^* & \frac{1}{\sqrt{2}}(\alpha_2 \alpha_4^*+\alpha_6 \alpha_5^*) & \frac{1}{2}(|\alpha_2|^2-|\alpha_6|^2) \\
 0 & \alpha_3 \alpha_1^* & \frac{1}{\sqrt{2}} \alpha_3 \alpha_2^*& |\alpha_3|^2 & \alpha_3 \alpha_4^* & \frac{1}{\sqrt{2}}\alpha_3 \alpha_2^* \\
 0 & \alpha_4 \alpha_1^* & \frac{1}{\sqrt{2}}(\alpha_4 \alpha_2^*+\alpha_5 \alpha_6^*) & \alpha_4 \alpha_3^* & |\alpha_4|^2+|\alpha_5|^2 &  \frac{1}{\sqrt{2}}(\alpha_4 \alpha_2^*-\alpha_5 \alpha_6^*) \\
 0 & \frac{1}{\sqrt{2}}\alpha_2 \alpha_1^* & 
 \frac{1}{2}(|\alpha_2|^2-|\alpha_6|^2) & 
 \frac{1}{\sqrt{2}}\alpha_2 \alpha_3^* & \frac{1}{\sqrt{2}}(\alpha_2 \alpha_4^*-\alpha_6 \alpha_5^*) & 
 \frac{1}{2}(|\alpha_2|^2+|\alpha_6|^2) \\
\end{array}
\right).
\end{equation}

The fact that new subspaces appear that prevent the state from fully decohere, entails more involved conditions for having long-lived entanglement.  
However, for our present purposes it suffices to focus, e.g., on the initial state with 
$\alpha_2=\alpha_3=\alpha_4=0$. In this case (\ref{rhoTinftADC}) reduces to
\begin{equation}\label{ADC2}
\rho(p\rightarrow 1)=\left(\begin{array}{cccccc}
0 & 0 & 0 & 0 & 0 & 0 \\
0 & |\alpha_1|^2 & 0 & 0 & 0 & 0 \\
0 & 0 & 0 & 0 & 0 & 0 \\
0 & 0 & 0 & 0 & 0 & 0 \\
0 & 0 & 0 & 0 & |\alpha_5|^2 & \alpha_5\alpha_6^* \\
0 & 0 & 0 & 0 & \alpha_5^*\alpha_6 & |\alpha_6|^2
\end{array}\right),
\end{equation}
whose corresponding concurrence reads
\begin{equation}
    C_f(p=1)=|\alpha_6|^2.
\end{equation}
In its turn, the asymptotic coherence, obtained from the representation (\ref{rhoTinftADC sl}), results
\beq\label{concuADC}
\mathcal{K}_{p=1}=|\alpha_6|^2+2\sqrt{2}|\alpha_6\alpha_5|.   
\eeq
This shows that indeed the fermionic concurrence can endure under dynamics other than the one considered in Sec. \ref{dinmodel}, and indicates that robust entanglement is not a feature distinctive of a specific type of evolution, but an attribute of dynamics in which persistent coherence subspaces exist. 

\section{Conclusions}
While systems of indistinguishable particles play a significant role in many quantum information tasks, there is still limited understanding of the dynamics of useful quantum resources in such systems, compared to the advances achieved in composites of distinguishable parties.
One of these appreciated quantum resources is entanglement, and therefore the strategies aimed at preserving it has become of primary importance. 
In particular, open systems are typically affected by decoherence, arising from  
the interaction with the environment and leading to the suppression of desirable quantum properties of the system, such as entanglement.
In this context, we contribute to the identification of strategies (here encoded in the preparation of appropriate initial states) that favor the preservation of entanglement in open systems of indistinguishable fermions subject to decoherence.

We investigated the non-dissipative Markovian evolution of entanglement and coherence in the simplest two-fermion system that exhibits fermionic entanglement (namely, two fermions with four accessible states each), collectively coupled to a thermal bosonic reservoir. 
States with persistent entanglement were identified,
demonstrating resilience to decoherence in a wide class of initial situations. 

Interestingly, the existence and amount of long-lived entanglement does not depend on the temperature of the bath.
Increasing $T$ leads to a faster loss of coherence and to a sooner arrival to the asymptotic concurrence, but does not play any role in the amount of enduring correlations.

From among all identified states with persistent entanglement, those that conform the family I stand out for possessing invariant, time-independent, entanglement throughout the evolution, while the density matrix and the coherence are evolving quantities. 
For such states the (constant) amount of fermionic concurrence depends only on the projections of the initial state on the Decoherence-Free subspace, spanned by the maximally entangled states  $|\psi_{3}^{-}\rangle=\tfrac{1}{\sqrt{2}}\left(\left|\psi_{14}^{sl}\right\rangle+\left|\psi_{23}^{sl}\right\rangle\right)$ and $|\psi_{6}^{-}\rangle=\tfrac{1}{\sqrt{2}}\left(\left|\psi_{14}^{sl}\right\rangle-\left|\psi_{23}^{sl}\right\rangle\right)$, which are the only vectors of the basis $\{|\psi_{n}^{-}\}$ that exhibit fermionic entanglement.

We extended the analysis to the (fermionic version of the) paradigmatic amplitude damping channel, and verified that persistent entanglement can also be achieved under such dissipative decoherence processes. 
We further observed that the decoherence-free subspace is not the only source of long-term coherence ---a necessary condition for having persistent entanglement---, since in general 
additional subspaces of long-lasting coherence can emerge. 
This indicates that fermionic concurrence can persist under a wider range of dynamics irrespective of its specific features, provided the evolution allows for the existence of subspaces in which a non-vanishing amount of coherence endures.

Since a non-zero amount of coherence is necessary for entanglement to exist, the preservation of fermionic concurrence implies the preservation of coherence (here in the Slater determinant basis).
Consequently, the identification of states with persistent entanglement also brings out families of states that do not fully decohere. 
Thus, in a broad sense, our 
results contribute to the identification of strategies aimed at preserving both quantum coherence and entanglement in open systems of indistinguishable fermions, an essential task for the optimal performance of numerous quantum technologies.

\ack{A. V. H. and E. S. were supported by DGAPA-UNAM through project PAPIIT IN112723. E. S. acknowledges financial support of
CONAHCyT (CVU 1229301).  A. P. M. acknowledge funding from Grants No. PICT 2020-SERIEA-00959 from ANPCyT (Argentina) and No. PIP 11220210100963CO from CONICET (Argentina) and partial support from SeCyT, Universidad Nacional de Córdoba (UNC), Argentina. }


\end{document}